\documentclass{article}
\usepackage{amsfonts}

\title{Schr\"odinger and related equations as Hamiltonian systems, manifolds of second-order tensors and new ideas of nonlinearity in quantum mechanics}

\author{J. J. S\l awianowski and V. Kovalchuk\\
Institute of Fundamental Technological Research,\\
Polish Academy of Sciences,\\
$5^{\rm B}$, Pawi\'{n}skiego str., 02-106 Warsaw, Poland\\
e-mails: jslawian@ippt.gov.pl, vkoval@ippt.gov.pl}

\begin{document}

\maketitle
\begin{abstract}
Considered is the Schr\"odinger equation in a finite-dimensional space as an equation of mathematical physics derivable from the variational principle and treatable in terms of the Lagrange-Hamilton formalism. It provides an interesting example of "mechanics" with singular Lagrangians, effectively treatable within the framework of Dirac formalism. We discuss also some modified "Schr\"odinger" equations involving second-order time derivatives and introduce a kind of non-direct, non-perturbative, geometrically-motivated nonlinearity based on making the scalar product a dynamical quantity. There are some reasons to expect that this might be a new way of describing open dynamical systems and explaining some quantum "paradoxes".
\end{abstract}


\tableofcontents

\section*{Introduction}

There was plenty of papers dealing with various aspects of the relationship between classical and quantum mechanics. The most popular topics are those concerning the quasiclassical asymptotics of quantum mechanics, the asymptotic expansions when $\hbar\rightarrow 0$ like the WKB approximation, oscillatory integrals and the method of stationary phase. There are also studies in the opposite direction, when, basing on the optical-mechanical analogy, the eikonal and Hamilton-Jacobi equations, one investigates purely classical structures having some striking analogies to quantum ones. And then one shows that really the mentioned classical structures, although a priori obvious on the purely classical level, may be also re-obtained from the quasiclassical limit transition $\hbar\rightarrow 0$ from quantum mechanics. Concerning such topics cf. for instance \cite{JJS_91} and first of all references therein. Other very interesting studies of this kind, based on the hydrodynamical picture of quantum mechanics, were presented by V. V. Kozlov \cite{Koz_03}. The author analyzed there vortices of the "quantum fluid" and that study is a part of his very interesting theory and methodology of vortices as a fundamental concept of dynamics. The ideas developed by this author go back to some very old and fundamental concepts in physics and philosophy of science.

Another very important branch of investigations was the comparative study of the classical and quantum dynamics, including the dynamics of open systems \cite{IKO_97,Jam_72,Jam_74}. Quantum measurement and decoherence problems in a sense belong to this topic \cite{BlJ_93_1,BlJ_93_2,BlJ_94,BlJ_96}. Certain ideas of nonlinearity in quantum mechanics appeared in connection to those problems (see, e.g., \cite{DoGo_96,DoGo_99,Gol_97,Svet_04,Svet_05} and references therein).

Our study, also as a matter of fact motivated by the aforementioned problems, is formally quite a different approach to the quantum-classical convolution of concepts. Namely, we "forget" what the Schr\"odinger equation physically is. For a moment, it is for us only a differential equation. To simplify the problem as far as possible, we consider a "finite-level system", when the "configuration space" is a finite set and the corresponding linear space of "wave functions" is finite-dimensional. All concepts we are here dealing with will remain essentially valid also in the infinite-dimensional case ("true" wave functions); obviously some care must be taken nevertheless when passing from a finite dimension to the infinite one.

Then we discuss the Lagrangian and Hamiltonian formalisms for such a "classical" mechanical system in $\mathbb{C}^{n}$. In particular, the Dirac theory of primary and secondary constraints for systems with degenerate Lagrangians is discussed. This approach enables one to formulate some models of nonlinearity. We hope this nonlinearity may be perhaps a tool for describing measurement paradoxes and decoherence. The main point is that our formalism seems to suggest in a natural way some geometrically well-justified nonlinearities, not ones introduced "by hand" as perturbations of some linear background. The main idea is that of dynamical "scalar product" which is not fixed once for all but itself is a dynamical quantity on equal footing with the "wave functions"; they both satisfy a closed system of essentially, non-perturbatively nonlinear differential equations. The structure of this nonlinearity is based and, one can say, almost canonically implied by the geometric structure of "classical" degrees of freedom. Because of this we hope that the resulting effectively nonlinear quantum mechanics may be perhaps free of paradoxes of decoherence and measurement and can provide some new description of open quantum systems, alternative to that described in \cite{BlJ_93_1,BlJ_93_2,BlJ_94,BlJ_96}.

Some ideas of our "classical" description of quantum systems in terms of phase spaces and Hamiltonian dynamics are similar to those suggested many years ago by D. Chru\'{s}ci\'{n}ski \cite{Chr_91}.

\section{Finite-level nonlinear Schr\"odinger equation in the Lagrange-Hamilton description}

Following the jargon used by laser specialists and those working with the quantum dynamics of mutually interacting spins, we use the term "finite-level quantum system" for such a one the "Hilbert space" of which is finite-dimensional, so it may be identified with $\mathbb{C}^{n}$, when some basis is fixed. However, we shall avoid the misuse of this identification, because it usually smuggles into the treatment some artificial objects obscuring and often just falsifying the proper geometric interpretation of the used concept and making impossible the introduction of new ideas.

And one thing must be explicitly stressed. We are in fact motivated by certain problems from the realm of foundations of quanta. However, for some reasons at this stage of our treatment it was convenient to pretend to "forget" about this motivation and just to consider Schr\"odinger equation in a finite-dimensional space purely classically, simply as an equation of mathematical physics, in a sense classical mechanics, derivable from the variational principle and because of this treatable in terms of Lagrange-Hamilton formalism. And it is really interesting even from the point of view of this "hypocritically" classical language. For instance, it provides an interesting example of "mechanics" with singular Lagrangians, effectively treatable within the framework of Dirac formalism involving the primary and secondary constraints in a phase space of the problem.

Later on we shall try to discuss the "Schr\"odinger" equation involving second-order time derivatives, and also introduce some kind of non-direct and geo\-metrically-motivated nonlinearity based on making the scalar product a dynamical quantity. There are some reasons to expect this might be a new way of describing open quantum systems and a new promising attempt towards explaining quantum "paradoxes", decoherence and measurement. And using the methods developed for $n$-level quantum systems, we formulate finally some ideas concerning the treatment in an infinite-dimensional Hilbert space and certain links with relativistic field equations.

\subsection{Some complex geometry in linear spaces}

In this section our primary concept is an $n(<\infty)$-dimensional linear space $W$ over the complex field $\mathbb{C}$. It is well known that such a space gives rise to the natural quadruple of mutually related $n$-dimensional complex linear spaces:
\begin{equation}\label{eq1}
W,\qquad W^{\ast},\qquad \overline{W},\qquad \overline{W}^{\ast}\simeq \overline{W^{\ast}},
\end{equation}
namely, $W$ itself, its usual dual $W^{\ast}$ over $\mathbb{C}$, i.e., the space of $\mathbb{C}$-linear $\mathbb{C}$-valued functions on $W$, and their complex-conjugate spaces $\overline{W}$ and $\overline{W}^{\ast}\simeq \overline{W^{\ast}}$. There are many mistakes and misunderstandings concerning the complex conjugate space $\overline{W}$ and the antidual one $\overline{W}^{\ast}\simeq \overline{W^{\ast}}$, which may be easily avoided at least if $W$ is finite-dimensional, that is the case at the present stage. Hence, $\overline{W}^{\ast}\simeq \overline{W^{\ast}}$ consists by definition of antilinear (semi-linear) functions on $W$, i.e., such ones which satisfy
\begin{equation}\label{eq2}
g(aw+bv)=\overline{a}g(w)+\overline{b}g(v)
\end{equation}
for arbitrary $w,v\in W$, $a,b\in \mathbb{C}$. So, there is a natural antilinear isomorphism of $W^{\ast}$ onto $\overline{W^{\ast}}$ given by
\begin{equation}\label{eq3}
W^{\ast}\ni f\mapsto \overline{f}\in \overline{W^{\ast}},\qquad \overline{f}(w):=\overline{f(w)}
\end{equation}
for any $w\in W$, $f\in W^{\ast}$; the complex conjugate is taken pointwisely. As mentioned, (\ref{eq3}) is an antilinear isomorphism acting between two linear spaces,
\begin{equation}\label{eq4}
\overline{(af+bg)}=\overline{a}\overline{f}+\overline{b}\overline{g}
\end{equation}
for any $f,g\in W^{\ast}$, $a,b\in \mathbb{C}$. For obvious reasons the inverse of (\ref{eq3}) will be denoted by the same symbol and the following holds:
\begin{equation}\label{eq5}
\overline{\overline{f}}=f.
\end{equation}

By analogy to the obvious canonical isomorphism between $W$ and $W^{\ast\ast}$, the complex-conjugate space $\overline{W}$ is defined as the antidual of $W^{\ast}$. Namely, its elements $u\in \overline{W}$ are by definition antilinear functions on $W^{\ast}$. So, for any $f\in W^{\ast}$, $u(f)$ is defined as
\begin{equation}\label{eq6}
u(f):=\overline{f(u)}=\overline{f}(u).
\end{equation}
Compare this with the standard identification of $w\in W$ with the linear function on $W^{\ast}$:
\begin{equation}\label{eq7}
w(f)=f(w).
\end{equation}
Again there exists a canonical antilinear isomorphism of $W$ onto $\overline{W}$:
\begin{equation}\label{eq8}
W\ni w\mapsto \overline{w}\in \overline{W},\qquad \overline{w}(f)=\overline{w(f)}.
\end{equation}

All this may be also done in other, equivalent way. The important things are the following ones. All linear spaces (\ref{eq1}) are logically different and the complex conjugates of vectors belong to other linear spaces, for example, $\overline{w}\in \overline{W}$ is not an element of $W$, unless some additional geometric structures are fixed in $W$. There exists canonical linear isomorphism between $W$ and $W^{\ast\ast}$ (and similarly between $\overline{W}$ and $\overline{W^{\ast\ast}}$). There exists also canonical antilinear isomorphism of $W$ and $\overline{W}$ (and similarly between $W^{\ast}$ and $\overline{W^{\ast}}$). But of course without additional geometric object in $W$ (some metric) there is neither any canonical isomorphism between $W$ and $W^{\ast}$ nor any one between $\overline{W}$ and $\overline{W^{\ast}}$.

If $(e_{1},\ldots,e_{n})$ and $(e^{1},\ldots,e^{n})$ are mutually dual bases in $W$ and $W^{\ast}$,
\begin{equation}\label{eq9}
e^{a}(e_{b})=\delta^{a}{}_{b},
\end{equation}
then by the complex conjugation one obtains from them canonically the mutually dual bases $(\overline{e}_{\bar{1}},\ldots,\overline{e}_{\overline{n}})$, $(\overline{e}^{\bar{1}},\ldots,\overline{e}^{\overline{n}})$ respectively in $\overline{W}$ and $\overline{W}^{\ast}\simeq \overline{W^{\ast}}$,
\begin{equation}\label{eq10}
\overline{e}^{\overline{a}}(\overline{e}_{\overline{b}})=
\delta^{\overline{a}}{}_{\overline{b}}.
\end{equation}
If we expand vectors and covectors in $W$ with respect to the first system of dual bases,
\begin{equation}\label{eq11}
w=w^{a}e_{a},\qquad f=f_{a}e^{a},\qquad f(w)=f_{a}w^{a},
\end{equation}
then, obviously,
\begin{equation}\label{eq12}
\overline{w}=\overline{w}^{\overline{a}}\overline{e}_{\overline{a}},\qquad \overline{f}=\overline{f}_{\overline{a}}\overline{e}^{\overline{a}}.
\end{equation}
In this special system of bases the complex conjugation of objects is analytically represented by the complex conjugate of the coefficients, as seen from the above formulae. But certainly $w$, $\overline{w}$, $f$, $\overline{f}$ are elements of pairwise different linear spaces. So, the above formulae introduce the system of notations which will be consequently used below.

Let us stress that the definition of $\overline{W^{\ast}}$ given above is valid in any linear space independently of its dimension being finite or not. Unlike this the definition of $\overline{W}$ quoted above works only in the algebraically reflexive linear ones, but no other ones present interest for us.

Further on, the whole system of tensor products of $W$, $W^{\ast}$, $\overline{W}$, $\overline{W^{\ast}}$ may be introduced. Obviously, in applications we are interested in, the most important are linear mappings of $W$ into itself, i.e., elements of
\begin{equation}\label{eq13}
{\rm L}(W)\simeq W\otimes W^{\ast}
\end{equation}
and sesquilinear forms, i.e., elements of $\overline{W^{\ast}}\otimes W^{\ast}$. The corresponding analytical matrix representation is given respectively by $L^{a}{}_{b}$, $F_{\overline{a}b}$ for $L\in {\rm L}(W)$, $F\in \overline{W^{\ast}}\otimes W^{\ast}$, etc. For sesquilinear forms we use the convention that they are antilinear in the first argument and linear in the second one:
\begin{equation}\label{eq14}
F(u,v)=F_{\overline{a}b}\overline{u}^{\overline{a}}v^{b},
\end{equation}
so that
\begin{equation}\label{eq15}
F(au+bw,v)=\overline{a}F(u,v)+\overline{b}F(w,v),\quad F(v,au+bw)=aF(v,u)+bF(v,w)
\end{equation}
for any values of vectors and coefficients. We usually need Hermitian forms, i.e., such ones that
\begin{equation}\label{eq16}
F(u,v)=\overline{F(v,u)},\qquad F_{\overline{a}b}=\overline{F}_{b\overline{a}}.
\end{equation}
Then, obviously,
\begin{equation}\label{eq17}
F(u,u)=\overline{F(u,u)},\qquad u\mapsto F(u,u)\in \mathbb{R}.
\end{equation}
If $F$ is non-degenerate,
\begin{equation}\label{eq18}
\det\left[F_{\overline{a}b}\right]\neq 0,
\end{equation}
i.e., if
\begin{equation}\label{eq19}
W\ni v\mapsto F(\cdot,v)\in \overline{W^{\ast}}
\end{equation}
is a linear isomorphism of $W$ onto $\overline{W^{\ast}}$, then there exists the inverse twice contravariant sesquilinear object $F^{-1}\in W\otimes \overline{W}$ with components denoted briefly by $F^{a\overline{b}}$ such that
\begin{equation}\label{eq20}
F^{a\overline{c}}F_{\overline{c}b}=\delta^{a}{}_{b},\qquad 
F_{\overline{a}c}F^{c\overline{b}}=\delta_{\overline{a}}{}^{\overline{b}}.
\end{equation}
Obviously, $\delta^{a}{}_{b}$, $\delta_{\overline{a}}{}^{\overline{b}}$ are here matrices of identity transformations respectively in $W$ and $\overline{W^{\ast}}$.

If $F$ is Hermitian and for any $u\in W$ we have
\begin{equation}\label{eq21}
F(u,u)>0\qquad {\rm if}\quad u\neq 0, 
\end{equation}
then we say that $F$ is positively definite. Then there exist such bases in $W$ that the matrix of $F$ is diagonal with $(+1)$-entries on the diagonal, 
\begin{equation}\label{eq22}
\left[F_{\overline{a}b}\right]=I_{n}.
\end{equation}
Otherwise $F$ has some signature (the number of positive and negative elements on the diagonal) which is an invariant of $F$.

Let us now assume that some non-degenerate Hermitian form $\Gamma\in \overline{W^{\ast}}\otimes W^{\ast}$ is fixed once for all in $W$, so we are dealing with the algebraic structure $(W,\Gamma)$. If $\Gamma$ is positively definite, $(W,\Gamma)$ is a finite-dimensional Hilbert space or, using more customary terms, a unitary space. If the positive definiteness is not assumed, we are dealing with a pseudounitary space. The form $\Gamma$ will play a role of the scalar product if we insist on the quantum-mechanical interpretation. So, it seems natural to expect it should be positively definite. However, in many formal problems this assumption is not necessary. Moreover, even within the framework of quantum-mechanical interpretation, one cannot exclude a priori models with non-definite scalar products, at least at some stage of considerations (remind the old quantization of electrodynamics according to the Gupta-Bleuler prescription).

\subsection{Lagrangian and Hamiltonian models of the usual and generalized Schr\"o\-dinger equations (evolution)}

Let us return to our "classical hypocrisy". We discuss some models of "analytical mechanics" in the "configuration space" $W$. The system has $2n$ degrees of freedom because the dimension of $W$ as a linear space over reals equals $2n$. However, often it is convenient to say that we are dealing with $n$ "complex degrees of freedom". If some basis $(e_{1},\ldots,e_{n})$ is fixed in $W$, then expanding the elements $\psi\in W$,
\begin{equation}\label{eq23}
\psi=\psi^{a}e_{a},
\end{equation}
we introduce "complex generalized coordinates" $\psi^{a}$. Their real and imaginary parts form a system of $2n$ usual real coordinates. It is more convenient to use another normalisation, namely,
\begin{equation}\label{eq24}
\psi^{a}=\frac{1}{\sqrt{2}}\left(x^{a}+i y^{a}\right),\qquad
\overline{\psi}^{\overline{a}}=\frac{1}{\sqrt{2}}\left(x^{a}-i y^{a}\right),
\end{equation}
where $x^{a},y^{a}\in \mathbb{R}$. Obviously, without any additional structure in $W$, the real and imaginary parts of the vector $\psi\in W$ are not well defined. Taking the real and imaginary parts of $\psi^{a}$ is an explicitly base-dependent procedure. Just like in certain well-known formulae of classical field theory it is often formally convenient to use the system of $2n$ complex variable $\psi^{a}$, $\overline{\psi}^{\overline{a}}$ as "coordinates". Many expressions and calculations look then simpler. Then $2n$ complex quantities $\psi^{a}$, $\overline{\psi}^{\overline{a}}$ are (on some intermediate stages) treated as if they were formally independent and only in the final formulae one "remembers" that they are interrelated. And expressions (\ref{eq24}) describe then something like an orthogonal transformation of the system of $2n$ real variables $(x^{a},y^{a})$ into the system of $2n$ complex variables $\left(\psi^{a},\overline{\psi}^{\overline{a}}\right)$. More precisely, in a sense we work in the complex space $W\times \overline{W}$ of the complex dimension $2n$. And later on the formulae are restricted to the "diagonal" of complex dimension $n$ (real dimension $2n$) consisting not of all pairs $(\psi,\varphi)\in W\times \overline{W}$ but only of the pairs of the form $\left(\psi,\overline{\psi}\right)\in W\times \overline{W}$. If there is some need of using symbols, we shall denote this "diagonal" as 
\begin{equation}\label{eq25}
{\rm Diag}\left(W\times \overline{W}\right):=\left\{\left(\psi,\overline{\psi}\right):\psi\in W\right\}.
\end{equation}
In variational problems one deals with real-valued Lagrangians. It is convenient to define them primarily as analytic functions on $W\times \overline{W}$ and perform all differential operations with respect to both arguments as independent ones. Of course, if non-constant, such functions are never real-valued. However, they are constructed so that to be real-valued on ${\rm Diag}\left(W\times \overline{W}\right)$. If they are expanded into double power series with respect to the $\left(W,\overline{W}\right)$-arguments, their coefficients must show Hermitian symmetry. To be more precise, this concerns functions of the potential energy type, i.e., which depend only on configurations. Lagrangians, however, depend on configurations (generalized coordinates) and velocities. So, they are functions on $W\times \overline{W}\times W\times \overline{W}\simeq W\times W\times \overline{W}\times \overline{W}$. Nevertheless the above statements concerning structure of potentials apply also to such tangent-bundle functions: simply $W$, $\overline{W}$ are then replaced by $W\times W$, $\overline{W\times W}\simeq \overline{W}\times \overline{W}$.

\subsubsection{Lagrangian for the standard Schr\"odinger equation}

Let us begin with the usual Schr\"odinger Lagrangian. Later on we shall consider the hierarchy of more and more complicated Lagrangians with the additional terms responsible for the various expected physical phenomena. 

\noindent{\bf Definition:}
The usual Schr\"odinger Lagrangian will be denoted by $L(1)$, where the label $(1)$ refers to the resulting first-order differential Euler-Lagrange equation. It is given by
\begin{equation}\label{eq26}
L(1)=i\alpha \Gamma_{\overline{a}b}\left(\overline{\psi}^{\overline{a}}\dot{\psi}^{b}
-\dot{\overline{\psi}}^{\overline{a}}\psi^{b}\right)
-\gamma\chi_{\overline{a}b}\overline{\psi}^{\overline{a}}\psi^{b},
\end{equation}
where $\alpha$, $\gamma$ are some constants, $\chi\in \overline{W^{\ast}}\otimes W^{\ast}$ is a Hermitian form on $W$, and obviously the dot-symbols are time derivatives.

It is interesting to mention that Lagrangians of this form appeared in a slightly different context in \cite{MoFe_53,San_82}. They were thought on as an alternative variational description of vibrations. And besides, the exchange of energy between real and imaginary parts of $\psi^{a}$ was expected to be some toy model of dissipative phenomena (the observed system and its unobserved partner). This has to do with so-called Birchhoffian formulation of dynamical laws.

Similar expressions appear also in Kozlov theory of vortices \cite{Koz_03} and in his study of optical, quantum-mechanical and quasiclassical phenomena. In particular, this concerns vortices appearing in hydrodynamical interpretation of the Schr\"odinger equation and its quasiclassical limit.

Due to the hermiticity of $\Gamma$ and $\chi$, $L(1)$ is real, i.e.,
\begin{equation}\label{eq27}
\overline{L(1)}=L(1),
\end{equation}
just as Lagrangian should be. The corresponding action is given by
\begin{equation}\label{eq28}
I(1)=\int L(1)dt
\end{equation}
and its variational derivative equals
\begin{equation}\label{eq29}
\frac{\delta I(1)}{\delta \overline{\psi}^{\overline{a}}(t)}=
\frac{\partial L(1)}{\partial \overline{\psi}^{\overline{a}}}-
\frac{d}{dt}\frac{\partial L(1)}{\partial \dot{\overline{\psi}}^{\overline{a}}}=
2i\alpha\Gamma_{\overline{a}b}\dot{\psi}^{b}-
\gamma\chi_{\overline{a}b}\psi^{b},
\end{equation}
where, according to the standard procedure mentioned above, $\psi$ and $\overline{\psi}$ are formally treated as independent quantities. Then, obviously, calculated in the dual way
\begin{equation}\label{eq30}
\frac{\delta I(1)}{\delta \psi^{a}(t)}=
\frac{\partial L(1)}{\partial \psi^{a}}-
\frac{d}{dt}\frac{\partial L(1)}{\partial \dot{\psi}^{a}}
\end{equation}
is the complex conjugate of (\ref{eq29}). Hence, it is only one of these expressions, by convention (\ref{eq29}), that is sufficient for obtaining equations of motion (Euler-Lagrange equations). Putting (\ref{eq29}) to vanish one obtains after some manipulations on tensor indices the following equation:
\begin{equation}\label{eq31}
i\alpha\frac{d\psi^{a}}{dt}=
\frac{\gamma}{2}\left({}^{\Gamma}\chi\right)^{a}{}_{b}\psi^{b},
\end{equation}
where ${}^{\Gamma}\chi=H\in {\rm L}_{\mathbb{C}}(W)$ is obtained from $\chi\in \overline{W^{\ast}}\otimes W^{\ast}$ by the $\Gamma$-raising of the first tensor index, i.e.,
\begin{equation}\label{eq32}
H^{a}{}_{b}=\left({}^{\Gamma}\chi\right)^{a}{}_{b}=\Gamma^{a\overline{c}}\chi_{\overline{c}b}.
\end{equation}
Obviously, $H={}^{\Gamma}\chi$ is by its very structure Hermitian with respect to $\Gamma$ ($\chi$ itself is objectively Hermitian),
\begin{equation}\label{eq33}
\Gamma\left(H\psi,\varphi\right)=
\Gamma\left(\psi,H\varphi\right).
\end{equation}

It is seen that (\ref{eq31}) becomes the literally understood Schr\"odinger equation when $\Gamma$ is positively definite, so in appropriate bases
\begin{equation}\label{eq34}
\left[\Gamma_{\overline{a}b}\right]=I_{n}
\end{equation}
and $\alpha=\hbar$, $\gamma=2$, where, obviously, $\hbar$ is the Planck constant.

Lagrangian (\ref{eq26}) is linear in generalized velocities $\dot{\psi}$, $\dot{\overline{\psi}}$ with coefficients depending linearly on generalized coordinates $\psi$, $\overline{\psi}$. Therefore, the action functional (\ref{eq28}) is quadratic with respect to the evolution curve $\mathbb{R}\ni t\mapsto \psi(t)\in W$ and its "stationarization" results in first-order linear differential equations for the time-dependence of $\psi$. As a matter of fact $\chi$, and then also $\widehat{H}$, may be time-dependent, and then the solution is not simply given by the corresponding operator exponent. In "usual" analytical mechanics Lagrangians linear in velocities and first-order differential equations of motion are rather exotic (although second-order differential equations may be in an obvious way reduced to doubled systems of first-order equations). And the resulting Legendre transformation is evidently non-invertible, one obtains constraints in the phase space of the system and the Dirac formalism of singular Hamiltonian mechanics must be used. As we shall see, this formalism for the Schr\"odinger equation has some geometric peculiarities and is interesting in itself.

\subsubsection{Admitting second derivatives}

If we continue to forget "hypocritically" about our quantum motivation, then from the point of view of purely classical analytical mechanics it is natural to ask after Lagrangians which lead to terms with second derivatives in equations of motion. 

\noindent{\bf Definition:}
Let us denote those Lagrangians by $L(1,2)$, where the labels $(1,2)$ refer to the occurrence of first and second time derivatives (velocities and accelerations) in equations of motion. Obviously, the simplest model is
\begin{equation}\label{eq35}
L(1,2)=i\alpha
\Gamma_{\overline{a}b}\left(\overline{\psi}^{\overline{a}}\dot{\psi}^{b}
-\dot{\overline{\psi}}^{\overline{a}}\psi^{b}\right)+
\beta\Gamma_{\overline{a}b}\dot{\overline{\psi}}^{\overline{a}}\dot{\psi}^{b}
-\gamma\chi_{\overline{a}b}\overline{\psi}^{\overline{a}}\psi^{b}.
\end{equation}

If $\alpha=0$, this becomes just the usual Lagrangian of the system of $2n$ coupled harmonic oscillators. However, being motivated by the quantum-mechanical problems and the usual Schr\"odinger equation, we are inclined to retain the $\alpha$-term; then the quadratic $\beta$-term is some kind of correction. Its physical interpretation in quantum-mechanical terms is still not clear, if possible at all.

\noindent{\bf Remark:}
One might formally admit to take another $\widetilde{\Gamma}$ in the term quadratic in velocities, then Lagrangian would have the following form: 
\begin{equation}\label{eq36}
\widetilde{L}(1,2)=\widetilde{L}\left(1,\Gamma;2,\widetilde{\Gamma}\right)=i\alpha
\Gamma_{\overline{a}b}\left(\overline{\psi}^{\overline{a}}\dot{\psi}^{b}
-\dot{\overline{\psi}}^{\overline{a}}\psi^{b}\right)+
\beta\widetilde{\Gamma}_{\overline{a}b}\dot{\overline{\psi}}^{\overline{a}}\dot{\psi}^{b}
-\gamma\chi_{\overline{a}b}\overline{\psi}^{\overline{a}}\psi^{b}.
\end{equation}
Of course, this makes the resulting equations much more complicated and one feels rather reluctant to such a modification, nevertheless it is formally possible. And perhaps it may be physically justified, provided of course that the quadratic correction may be physically interpretable at all.

Let us observe that from the point of view of purely classical analytical mechanics the quantities $\Gamma$, $\widetilde{\Gamma}$ are logically independent and it is fully justified to discuss dynamical models in which they are different and non-correlated to each other. Obviously, the resulting equations of motion would be then more complicated and the symmetry group rather restricted (because it must preserve two different sesquilinear forms $\Gamma$, $\widetilde{\Gamma}$). The mathematical peculiarity of $\Gamma$, $\widetilde{\Gamma}$ proportional to each other is just the "large" symmetry group of equations of motion.

Nevertheless, our analysis has some "quantum" motivation, where $\Gamma$ is to play the role of scalar product (or some generalised scalar product). And then it is more reasonable and physically justified to use only one form $\Gamma$.

\noindent{\bf Remark:} 
Structural similarity of Lagrangian (\ref{eq26}) to that underlying Dirac equation is obvious, although Dirac equation is a partial one, and here for simplicity we deal with ordinary differential equations (finite "configuration space"). This similarity is not accidental (see e.g. our papers \cite{JJS_96,JJS_98,JJS_01,KGD_02}). The same concerns (\ref{eq35}) where in the resulting differential equations second-order time derivatives are combined with first-order ones \cite{JMK_01,JMK_03}.

The corresponding action functionals will be denoted respectively by
\begin{equation}\label{eq37}
I(1,2)=\int L(1,2)dt,\qquad \widetilde{I}(1,2)=\int \widetilde{L}(1,2)dt.
\end{equation}
Putting $\alpha=0$ we obtain the "usual" Lagrangian of analytical mechanics (for $2n$ coupled harmonic oscillators, as said above). The corresponding Lagrangian will be denoted by $L(2)$,
\begin{equation}\label{eq38}
L(2)=\beta\Gamma_{\overline{a}b}
\dot{\overline{\psi}}^{\overline{a}}\dot{\psi}^{b}
-\gamma\chi_{\overline{a}b}\overline{\psi}^{\overline{a}}\psi^{b},
\end{equation}
and its action functional by $I(2)$,
\begin{equation}\label{eq39}
I(2)=\int L(2)dt.
\end{equation}
It is obvious that 
\begin{equation}\label{eq40}
\frac{\delta I(1,2)}{\delta \overline{\psi}^{\overline{a}}(t)}=
2i\alpha\Gamma_{\overline{a}b}\frac{d\psi^{b}}{dt}-
\beta\Gamma_{\overline{a}b}\frac{d^{2}\psi^{b}}{dt^{2}}-
\gamma\chi_{\overline{a}b}\psi^{b}.
\end{equation}
There is a qualitative catastrophic discontinuity if performing the limit transition $\beta\rightarrow 0$ in the corresponding equations of motion:
\begin{equation}\label{eq41}
i\alpha\frac{d\psi^{a}}{dt}-\frac{\beta}{2}\frac{d^{2}\psi^{a}}{dt^{2}}=
\frac{\gamma}{2}H^{a}{}_{b}\psi^{b},
\end{equation}
where the "quantum" Hamiltonian $H^{a}{}_{b}$ is given again by (\ref{eq32}). This is nothing else but the special case of the general phenomenon that differential equations and phase portraits of the corresponding dynamical systems are drastically unstable with respect to neglecting the highest-order derivative terms.

Obviously, on the level of pure analytical mechanics it is just (\ref{eq38}) and (\ref{eq39}), i.e., $\alpha=0$ situation, that is the most natural model. If we do not hide the (true) quantum-mechanical motivation, then obviously the model (\ref{eq26}), i.e., $\beta=0$ situation, seems to be just the true model. The natural question arises as to the status of the second derivatives. Certain arguments were raised by various authors in favour of the $\beta$-term in $L(1,2)$, i.e., "quantum mechanics" corrected by the second-derivative ("acceleration") expression in the equation of evolution for the "wave function" $\psi$ \cite{Dvoe_05,GF_04,Krug_04,KGD_02}. It does not seem yet clear, however, if this term may be made compatible with the statistical interpretation. And if it is incompatible, what might be a possible alternative interpretation to be used instead. This finite-level problem resembles the interplay of first- and second-order derivatives of wave amplitudes in Dirac and Klein-Gordon relativistic equations. And it is really a "discretized" version of the problem. Moreover, the mixing of terms like in (\ref{eq35}) and (\ref{eq41}) is known from the field-theoretic treatments. There exist relativistic models in which Dirac and d'Alembert operators are superposed \cite{Dvoe_05,GF_04,Krug_04,JJS_96,JJS_98,JJS_01,KGD_02}. There were various motivations for that. One of them was the demand of conformal invariance and the gauge model of gravitation based on the conformal group \cite{GF_04,JJS_96,JJS_98,JJS_01,KGD_02}. Incidentally, the mentioned models have some quite unexpected and interesting consequences which may be of some relevance for the standard model.

We do not discuss here the model (\ref{eq36}) with different Hermitian metrics $\Gamma$, $\widetilde{\Gamma}$ for the first- and second-order terms. Nevertheless, let us quote the expression for its variational derivative and the corresponding "Schr\"odinger equation" in the form solved with respect to the first derivatives:
\begin{eqnarray}
\frac{\delta \widetilde{I}(1,2)}{\delta \overline{\psi}^{\overline{a}}(t)}&=&
2i\alpha\Gamma_{\overline{a}b}\frac{d\psi^{b}}{dt}-
\beta\widetilde{\Gamma}_{\overline{a}b}\frac{d^{2}\psi^{b}}{dt^{2}}-
\gamma\chi_{\overline{a}b}\psi^{b},\label{eq42}\\
2i\alpha\frac{d\psi^{a}}{dt}&=&
\beta\left({}^{\Gamma}\widetilde{\Gamma}\right)^{a}{}_{b}
\frac{d^{2}\psi^{b}}{dt^{2}}+
\gamma H^{a}{}_{b}\psi^{b},\label{eq43}
\end{eqnarray}
where ${}^{\Gamma}\widetilde{\Gamma}\in {\rm L}(W)$ is given by
\begin{equation}\label{eq44}
\left({}^{\Gamma}\widetilde{\Gamma}\right)^{a}{}_{b}:=
\Gamma^{a\overline{c}}\widetilde{\Gamma}_{\overline{c}b}
\end{equation}
and $H={}^{\Gamma}\chi\in {\rm L}(W)\simeq W\otimes W^{\ast}$ is, as usual, the "Hamiltonian" (\ref{eq32}).

\subsection{Admitting "potentials" and direct nonlinearity}

The above models are linear and finite-dimensional with perhaps some time-dependent coefficients ($\chi$, or equivalently $H$). Therefore, they are formally equivalent to a finite system of harmonic oscillators with a possible parametric-type excitation. Incidentally, to include non-parametric-type excitation, one has to admit Lagrangians to be general second-order polynomials of the state variables, not necessarily quadratic forms. The terms like
\begin{equation}\label{eq45}
L_{\rm exc}=F_{a}\psi^{a}+
\overline{F}_{\overline{a}}\overline{\psi}^{\overline{a}}
\end{equation}
with time-dependent $F_{a}$ describe in analytical mechanics the extra-imposed external extortion. The resulting Euler-Lagrange equations are affine ("linear-non-homogeneous") but no longer literally linear in state variables. So, although natural and very well known in analytical mechanics, they are outside the scope of quantum mechanics with its linear Schr\"odinger equation. But being once faced with such an "elementary" nonlinearity ("pseudo-nonlinearity", so to speak, in the sense "linear-non-homogeneous" or "affine"), one feels motivated to admit a real, serious nonlinearity by introducing to $L$ some general potential. In analytical mechanics this is something very natural and belonging to the everyday practice. The "usual" quantum mechanics is linear and within this framework such corrections might seem exotic. Nevertheless, it is well known that there exists some motivation and there were some attempts to introduce nonlinearity to quantum mechanics. They have to do with so-called "paradoxes" like the reduction of wave packet, decoherence, measurement, etc. Many of such attempts were "blind" in the sense of introducing nonlinearity "by hand" using the trial-and-error methodology. The language of geometric models in analytical mechanics enables one to proceed in a more systematic way basing on some natural guiding hints.

As mentioned, the simplest way is to introduce to Lagrangian some potential term $V$ built in a non-quadratic way of the "wave function" $\psi$. This is still a "traditional" way based on some kind of ad hoc "experiments" with postulated potentials. The method we suggest in this paper, namely the one based on the dynamical scalar product, seems much more natural and promising. Nevertheless, let us mention in a few words the traditional procedure. 

\noindent{\bf Definition:} 
The corresponding Lagrangians will be denoted by $L({\rm dnl})$, where "dnl" means "directly nonlinear". They may be based either on the Schr\"odinger linear background $L(1)$ (\ref{eq26}) or on the linear background $L(1,2)$ (\ref{eq35}) predicting the second-derivative term,
\begin{eqnarray}
L(1,{\rm dnl})&=&L(1)+V,\label{eq46}\\
L(1,2,{\rm dnl})&=&L(1,2)+V,\label{eq47}
\end{eqnarray}
where $V$ is just the potential term built in a non-quadratic way of $\psi$ and responsible for the "direct nonlinearity".

The corresponding contribution to the action functional will be denoted by
\begin{equation}\label{eq48}
I(V):=\int Vdt.
\end{equation}
If we use the complex formalism, it is convenient to follow the procedure described above. Namely, we start from some analytic function of two variables, i.e., $V(\psi,\varphi)$, $\psi\in W$, $\varphi\in \overline{W}$, satisfying the aforementioned hermiticity condition, therefore real on the diagonal $\left\{\left(\psi,\overline{\psi}\right):\psi\in W\right\}$.

\noindent{\bf Proposition:} 
The most natural expressions are ones built of the obvious invariant $\Gamma_{\overline{a}b}\overline{\psi}^{\overline{a}}\psi^{b}$,
\begin{equation}\label{eq49}
V\left(\psi,\overline{\psi}\right)=
f\left(\Gamma_{\overline{a}b}\overline{\psi}^{\overline{a}}\psi^{b}\right),
\end{equation}
where $f:\mathbb{R}\rightarrow\mathbb{R}$ is some real-valued function of one real variable.

For example, one can think about the "quartic" model often used in quantum field theory and elementary particles physics:
\begin{equation}\label{eq50}
f(x)=\kappa(x-a)^{2},
\end{equation}
where $\kappa,a\in\mathbb{R}$ denote some real constants. Obviously, the variational derivative of the corresponding functional $I(V)$ is given by
\begin{equation}\label{eq51}
\frac{\delta I(V)}{\delta\overline{\psi}^{\overline{a}}(t)}=f^{\prime}
\left(\Gamma_{\overline{c}d}\overline{\psi}^{\overline{c}}\psi^{d}\right)
\Gamma_{\overline{a}b}\psi^{b},
\end{equation}
where $f^{\prime}$ is the usual first-order derivative of $f:\mathbb{R}\rightarrow\mathbb{R}$.

Variation with respect to $\psi^{a}$ is given by the complex-conjugate expression:
\begin{equation}\label{eq52}
\frac{\delta I(V)}{\delta\psi^{a}(t)}=f^{\prime}
\left(\Gamma_{\overline{c}d}\overline{\psi}^{\overline{c}}\psi^{d}\right)
\overline{\psi}^{\overline{b}}\Gamma_{\overline{b}a}.
\end{equation}
The corresponding nonlinear Schr\"odinger equation with the possible second-order differential term has the following form:
\begin{equation}\label{eq53}
2i\alpha\Gamma_{\overline{a}b}\frac{d\psi^{b}}{dt}-
\beta\Gamma_{\overline{a}b}\frac{d^{2}\psi^{b}}{dt^{2}}=
\gamma\chi_{\overline{a}b}\psi^{b}+f^{\prime}\Gamma_{\overline{a}b}\psi^{b},
\end{equation}
i.e.,
\begin{equation}\label{eq54}
i\alpha\frac{d\psi^{a}}{dt}-\frac{\beta}{2}\frac{d^{2}\psi^{a}}{dt^{2}}=
\frac{\gamma}{2}H^{a}{}_{b}\psi^{b}+\frac{1}{2}f^{\prime}\psi^{a}.
\end{equation}

\subsection{Canonical formalism}

We shall now discuss some problems of canonical formalism for the above usual and modified Schr\"odinger equations. Before doing this we again return to some comments concerning our complex language. As mentioned, just like in some studies concerning classical field theory and its quantization, it is convenient to use the complex formalism. Let us adapt it to the phase-space description. As mentioned, in a fixed basis in $W$ we put
\begin{equation}\label{eq55}
\psi^{a}=\frac{1}{\sqrt{2}}\left(x^{a}+iy^{a}\right),\qquad
\overline{\psi}^{\overline{a}}=\frac{1}{\sqrt{2}}\left(x^{a}-iy^{a}\right),
\end{equation}
where $x^{a}$, $y^{a}$ form a system of $2n$ real coordinates in the configuration space. In other words, we perform analytical continuation from the real form of $W$ to the complex space $W\times \overline{W}$ and then perform the restriction to the diagonal ${\rm Diag}\left(W\times \overline{W}\right)$ (\ref{eq25}). If we use the language of real geometry, the canonical momenta conjugate to $\left(x^{a},y^{a}\right)$ are denoted respectively by $\left(u_{a},v_{a}\right)$. They are coordinates in the dual space $W^{\ast}$ (as a real space). In the language of complex geometry, we use the momentum space $W^{\ast}\times \overline{W}^{\ast}$, i.e., the phase space $W\times \overline{W}\times W^{\ast}\times \overline{W}^{\ast}$. The momenta conjugate to $\left(\psi^{a},\overline{\psi}^{\overline{a}}\right)$ are denoted by $\left(\pi_{a},\overline{\pi}_{\overline{a}}\right)$, where obviously
\begin{equation}\label{eq56}
\pi_{a}=\frac{1}{\sqrt{2}}\left(u_{a}-iv_{a}\right),\qquad
\overline{\pi}_{\overline{a}}=\frac{1}{\sqrt{2}}\left(u_{a}+iv_{a}\right).
\end{equation}
Inverting formulae (\ref{eq55}), (\ref{eq56}) we obtain
\begin{eqnarray}
x^{a}=\frac{1}{\sqrt{2}}\left(\psi^{a}+\overline{\psi}^{\overline{a}}\right),&\quad&
y^{a}=-\frac{i}{\sqrt{2}}\left(\psi^{a}-\overline{\psi}^{\overline{a}}\right),
\label{eq57a}\\
u_{a}=\frac{1}{\sqrt{2}}\left(\pi_{a}+\overline{\pi}_{\overline{a}}\right),&\quad&
v_{a}=\frac{i}{\sqrt{2}}\left(\pi_{a}-\overline{\pi}_{\overline{a}}\right).
\label{eq57b}
\end{eqnarray}
Formally we are dealing here with the orthogonal change of variables from $(x,y)$ to $\left(\psi,\overline{\psi}\right)$ with the block matrix
\begin{equation}\label{eq58}
\frac{1}{\sqrt{2}}\left[
\begin{array}{cc}
I_{n} & iI_{n} \\
I_{n} & -iI_{n}
\end{array}
\right],
\end{equation}
where $I_{n}$ obviously denotes the $n\times n$ identity matrix. Transformation from $(u,v)$ to $\left(\pi,\overline{\pi}\right)$ is obviously given by the contragradient (inverse and transposed) matrix. The relationship between differential operators in $(x,y;u,v)$ and the corresponding analytical continuation to $\left(\psi,\overline{\psi};\pi,\overline{\pi}\right)$ is evidently given by
\begin{eqnarray}
\frac{\partial}{\partial x^{a}}=
\frac{1}{\sqrt{2}}\frac{\partial}{\partial \psi^{a}}+
\frac{1}{\sqrt{2}}\frac{\partial}{\partial \overline{\psi}^{\overline{a}}},&\quad&
\frac{\partial}{\partial y^{a}}=
\frac{i}{\sqrt{2}}\frac{\partial}{\partial \psi^{a}}-
\frac{i}{\sqrt{2}}\frac{\partial}{\partial \overline{\psi}^{\overline{a}}},
\label{eq59a}\\
\frac{\partial}{\partial u_{a}}=
\frac{1}{\sqrt{2}}\frac{\partial}{\partial \pi_{a}}+
\frac{1}{\sqrt{2}}\frac{\partial}{\partial \overline{\pi}_{\overline{a}}},&\quad&
\frac{\partial}{\partial v_{a}}=
-\frac{i}{\sqrt{2}}\frac{\partial}{\partial \pi_{a}}+
\frac{i}{\sqrt{2}}\frac{\partial}{\partial \overline{\pi}_{\overline{a}}},
\label{eq59b}
\end{eqnarray}
and conversely
\begin{eqnarray}
\frac{\partial}{\partial \psi^{a}}=
\frac{1}{\sqrt{2}}\frac{\partial}{\partial x^{a}}-
\frac{i}{\sqrt{2}}\frac{\partial}{\partial y^{a}},&\quad&
\frac{\partial}{\partial \overline{\psi}^{\overline{a}}}=
\frac{1}{\sqrt{2}}\frac{\partial}{\partial x^{a}}+
\frac{i}{\sqrt{2}}\frac{\partial}{\partial y^{a}},
\label{eq60a}\\
\frac{\partial}{\partial \pi_{a}}=
\frac{1}{\sqrt{2}}\frac{\partial}{\partial u_{a}}+
\frac{i}{\sqrt{2}}\frac{\partial}{\partial v_{a}},&\quad&
\frac{\partial}{\partial \overline{\pi}_{\overline{a}}}=
\frac{1}{\sqrt{2}}\frac{\partial}{\partial u_{a}}-
\frac{i}{\sqrt{2}}\frac{\partial}{\partial v_{a}}.
\label{eq60b}
\end{eqnarray}
The symplectic form 
\begin{equation}\label{eq61}
\gamma=du_{a}\wedge dx^{a} + dv_{a}\wedge dy^{a},
\end{equation}
after analytical continuation looks as follows:
\begin{equation}\label{eq62}
\gamma=d\pi_{a}\wedge d\psi^{a} + d\overline{\pi}_{\overline{a}}\wedge d\overline{\psi}^{\overline{a}},
\end{equation}
and, as expected, the Poisson brackets
\begin{equation}\label{eq63}
\{f,g\}=\frac{\partial f}{\partial x^{a}}\frac{\partial g}{\partial u_{a}}+
\frac{\partial f}{\partial y^{a}}\frac{\partial g}{\partial v_{a}}-
\frac{\partial f}{\partial u_{a}}\frac{\partial g}{\partial x^{a}}-
\frac{\partial f}{\partial v_{a}}\frac{\partial g}{\partial y^{a}}
\end{equation}
after analytical continuation become
\begin{equation}\label{eq64}
\{f,g\}=\frac{\partial f}{\partial \psi^{a}}\frac{\partial g}{\partial \pi_{a}}+
\frac{\partial f}{\partial \overline{\psi}^{\overline{a}}}\frac{\partial g}{\partial \overline{\pi}_{\overline{a}}}-
\frac{\partial f}{\partial \pi_{a}}\frac{\partial g}{\partial \psi^{a}}-
\frac{\partial f}{\partial \overline{\pi}_{\overline{a}}}\frac{\partial g}{\partial \overline{\psi}^{\overline{a}}}.
\end{equation}
Hamiltonian vector fields with generators $F$,
\begin{equation}\label{eq65}
X_{F}=\frac{\partial F}{\partial u_{a}}\frac{\partial}{\partial x^{a}}+
\frac{\partial F}{\partial v_{a}}\frac{\partial}{\partial y^{a}}-
\frac{\partial F}{\partial x^{a}}\frac{\partial}{\partial u_{a}}-
\frac{\partial F}{\partial y^{a}}\frac{\partial}{\partial v_{a}},
\end{equation}
become obviously
\begin{equation}\label{eq66}
X_{F}=\frac{\partial F}{\partial \pi_{a}}\frac{\partial}{\partial \psi^{a}}+
\frac{\partial F}{\partial \overline{\pi}_{\overline{a}}}\frac{\partial}{\partial \overline{\psi}^{\overline{a}}}-
\frac{\partial F}{\partial \psi^{a}}\frac{\partial}{\partial \pi_{a}}-
\frac{\partial F}{\partial \overline{\psi}^{\overline{a}}}\frac{\partial}{\partial \overline{\pi}_{\overline{a}}}.
\end{equation}

Let us quote some additional obvious formulae which often appear in our calculus concerning the above and other formulae:
\begin{eqnarray}
\left\langle d\psi^{a},\frac{\partial}{\partial \psi^{b}}\right\rangle =
\delta^{a}{}_{b},&\quad&
\left\langle d\psi^{a},\frac{\partial}{\partial \overline{\psi}^{\overline{b}}}\right\rangle = 0,
\label{eq67a}\\
\left\langle d\overline{\psi}^{\overline{a}},\frac{\partial}{\partial \psi^{b}}\right\rangle = 0,&\quad&
\left\langle d\overline{\psi}^{\overline{a}},\frac{\partial}{\partial \overline{\psi}^{\overline{b}}}\right\rangle = \delta^{\overline{a}}{}_{\overline{b}}.
\label{eq67b}
\end{eqnarray}
Similarly,
\begin{eqnarray}
\left\langle d\psi^{a},\frac{\partial}{\partial \pi_{b}}\right\rangle = 0,&\quad&
\left\langle d\psi^{a},\frac{\partial}{\partial \overline{\pi}_{\overline{b}}}\right\rangle = 0,
\label{eq68a}\\
\left\langle d\pi_{a},\frac{\partial}{\partial \pi_{b}}\right\rangle = \delta_{a}{}^{b},&\quad&
\left\langle d\pi_{a},\frac{\partial}{\partial \psi^{b}}\right\rangle = 0,\qquad \ldots
\label{eq68b}
\end{eqnarray}
Concerning the formulae like (\ref{eq62}), (\ref{eq64}), (\ref{eq66}), and so on, it must be stated that working in the "configuration space" $W\times \overline{W}$ and the "phase space" $W\times \overline{W}\times W^{\ast}\times \overline{W}^{\ast}$ is an auxiliary tool, although very convenient one. However, the true "physical" phenomena take place in $W$ and $W\times W^{\ast}$ as the configuration and phase spaces, respectively. Fortunately, the analytical continuation from ${\rm Diag}\left(W\times \overline{W}\right)$ enables one to work almost automatically in the mentioned spaces of the doubled dimension.

\subsubsection{Legendre transformation, constraints, Dirac procedure}

Let us consider the Legendre transformation based on the usual Schr\"odinger Lagrangian (\ref{eq26}). As the corresponding action functional $\psi\mapsto I(1)(\psi)$ is quad\-ratic and the resulting "equations of motion" are linear, this study is rather academic. Nevertheless, it remains essentially valid when some direct nonlinearity is introduced by the "potential energy" term $V\left(\psi,\overline{\psi}\right)$ and everything becomes especially instructive when the term quadratic in velocities is introduced, i.e., if we consider the model $L(1,2)$ (\ref{eq35}) and the problem of the limit transition $\beta\rightarrow 0$, a very singular one. For the pure model $L(1)$ the Legendre transformation has the following form:
\begin{equation}\label{eq69}
\pi_{a}=i\alpha\Gamma_{\overline{b}a}\overline{\psi}^{\overline{b}},\qquad 
\overline{\pi}_{\overline{a}}=-i\alpha\Gamma_{\overline{a}b}\psi^{b},
\end{equation}
the second equation being the complex conjugate of the first one; this is the obvious redundancy following from the use of complex language. It is seen that canonical momenta are completely independent of generalized velocities, so this is an extreme case of singular Dirac mechanics (compare the situation, by the way, with one for the relativistic Dirac equation). Therefore, the primary constraints are redundantly described by equations
\begin{equation}\label{eq70}
\phi_{a}=0,\qquad \overline{\phi}_{\overline{a}}=0,
\end{equation}
where
\begin{equation}\label{eq71}
\phi_{a}=\pi_{a}-i\alpha\Gamma_{\overline{b}a}\overline{\psi}^{\overline{b}},
\qquad \overline{\phi}_{\overline{a}}=\overline{\pi}_{\overline{a}}+
i\alpha\Gamma_{\overline{a}b}\psi^{b}.
\end{equation}
According to the general rules of Lagrangian-Hamiltonian mechanics, the "energy" function $e$ is given by
\begin{equation}\label{eq72}
e=\dot{\psi}^{a}\frac{\partial L}{\partial \dot{\psi}^{a}}+
\dot{\overline{\psi}}^{\overline{a}}\frac{\partial L}{\partial \dot{\overline{\psi}}^{\overline{a}}}-L,
\end{equation}
in this case $L=L(1)$ (\ref{eq26}). Of course, one should not confuse this "energy" in the sense of Hamiltonian mechanics with quantum energy given by the Hamilton operator (\ref{eq32}). After some trivial calculations we obtain that
\begin{equation}\label{eq73}
e=\gamma\chi_{\overline{a}b}\overline{\psi}^{\overline{a}}\psi^{b},
\end{equation}
so the "energy" function $e$ is independent of generalized velocities --- a rather exotic property. As usually, $e$ is a pull-back, under Legendre transformation $\mathcal{L}$, of some "Hamiltonian" $h$ defined only on the manifold 
\begin{equation}\label{eq73a}
M=\mathcal{L}\left(W\times \overline{W}\times W^{\ast}\times \overline{W}^{\ast}\right)
\end{equation}
of primary Dirac constraints in the phase space of the system. However, according to the traditional procedure of Dirac, it is convenient to use the family of "Hamiltonians" $\mathcal{H}$ defined on the "total phase space" $W\times \overline{W}\times W^{\ast}\times \overline{W}^{\ast}$ and having the property that
\begin{equation}\label{eq74}
\mathcal{H}|M=h.
\end{equation}
The standard procedure is to fix any such a Hamiltonian $\mathcal{H}_{0}$ and then to put
\begin{equation}\label{eq75}
\mathcal{H}=\mathcal{H}_{0}+\lambda^{a}\phi_{a}+
\overline{\lambda}^{\overline{a}}\overline{\phi}_{\overline{a}}
\end{equation}
with yet undetermined Lagrange factors $\lambda^{a}$, $\overline{\lambda}^{\overline{a}}$. The most natural, almost canonical choice is
\begin{equation}\label{eq76}
\mathcal{H}_{0}=
\gamma\chi_{\overline{a}b}\overline{\psi}^{\overline{a}}\psi^{b}.
\end{equation}
Then, following the well-known Dirac procedure, one must determine the submanifold $M_{s}$ of secondary constraints, eliminate as fast as possible the above Lagrange coefficients $\lambda^{a}$, $\overline{\lambda}^{\overline{a}}$ ("gauge variables"), define the effective Hamiltonian on $M_{s}$ and the corresponding Poisson brackets of functions on $M_{s}$ (Dirac brackets in a sense). The first step is to calculate Hamiltonian vector fields $X_{\mathcal{H}}$,
\begin{equation}\label{eq77}
X_{\mathcal{H}}=X_{\mathcal{H}_{0}}+X_{\lambda^{a}\phi_{a}}+
X_{\overline{\lambda}^{\overline{a}}\overline{\phi}_{\overline{a}}}.
\end{equation}
More precisely, it is sufficient to take the simplified form of $X_{\mathcal{H}}$, namely,
\begin{equation}\label{eq78}
X_{\mathcal{H}}=X_{\mathcal{H}_{0}}+\lambda^{a}X_{\phi_{a}}+
\overline{\lambda}^{\overline{a}}X_{\overline{\phi}_{\overline{a}}},
\end{equation}
because the expressions (\ref{eq77}), (\ref{eq78}) evidently coincide on the manifold of primary constraints. Then, the care must be taken to make $X_{\mathcal{H}}$ compatible with constraints $M$, i.e., being tangent to it. At points of non-tangency the dynamics is inconsistent.

It is a trivial task to calculate the contractions of differentials of (\ref{eq70}), (\ref{eq71}) with the vector fields (\ref{eq77}), (\ref{eq78}). As differentials of constraints equations are given in the redundant space $W\times \overline{W}\times W^{\ast}\times \overline{W}^{\ast}$ by
\begin{equation}\label{eq79}
d\phi_{c}=d\pi_{c}-i\alpha\Gamma_{\overline{d}c}d\overline{\psi}^{\overline{d}}, \qquad d\overline{\phi}_{\overline{c}}=d\overline{\pi}_{\overline{c}}+
i\alpha\Gamma_{\overline{c}d}d\psi^{d},
\end{equation}
the tangency conditions
\begin{equation}\label{eq80}
\left\langle d\phi_{c},X_{\mathcal{H}}\right\rangle = 0,\qquad
\left\langle d\overline{\phi}_{\overline{c}},X_{\mathcal{H}}\right\rangle = 0
\end{equation}
have obviously the unique solutions for $\lambda^{a}$, $\overline{\lambda}^{\overline{a}}$, i.e.,
\begin{equation}\label{eq81}
\lambda^{a}=-\frac{i}{2}\frac{\gamma}{\alpha}\Gamma^{a\overline{c}}
\chi_{\overline{c}b}\psi^{b},\qquad \overline{\lambda}^{\overline{a}}=
\frac{i}{2}\frac{\gamma}{\alpha}\overline{\psi}^{\overline{b}}
\chi_{\overline{b}c}\Gamma^{c\overline{a}},
\end{equation}
all over the primary constraints manifold $M$. No additional restrictions are imposed on the admissible points of $M$ and because of this the secondary constraints $M_{s}$ (in the Dirac sense) coincide with the manifold of primary constraints. No gauge freedom appears because at all points of $M=M_{s}$ the Lagrange multipliers are uniquely defined by (\ref{eq81}). The momentum variables $\pi_{a}$, $\overline{\pi}_{\overline{a}}$ are uniquely determined by the generalized coordinates $\psi^{a}$, $\overline{\psi}^{\overline{a}}$. Let $i_{M}=i_{M_{s}}$ denote the natural injection of $M=M_{s}$ into the phase space manifold, then
\begin{equation}\label{eq82}
\gamma||M=\gamma||M_{s}=i_{M}^{\ast}\gamma=i_{M_{s}}^{\ast}\gamma,
\end{equation}
i.e., the restriction of the natural symplectic two-form on the primary phase space $P$ to $M=M_{s}$ is simply given by
\begin{equation}\label{eq83}
\gamma||M=\gamma||M_{s}=2i\alpha\Gamma_{\overline{a}b}
d\overline{\psi}^{\overline{a}}\wedge d\psi^{b}.
\end{equation}
Roughly speaking, this means that on $M=M_{s}$ the quantities $\overline{\psi}^{\overline{a}}$ become effectively the canonical momenta conjugate to $\psi^{a}$ as the generalized coordinates (obviously, up to the linear transformation with matrix $2i\alpha\Gamma_{\overline{a}b}$). 

Some subtle points appear here due to the use of complex coordinates $\psi^{a}$. Since their number is arbitrary, it takes any admissible value $n$, not necessarily an even one as it must be in symplectic manifolds. But $M=M_{s}$ has the complex dimension $n$, whereas its real one is always even and equals $2n$, cf. the formulae (\ref{eq55})--(\ref{eq57b}), (\ref{eq59a})--(\ref{eq62}). Therefore, the $n$ complex parameters
\begin{equation}\label{eq84}
\Pi_{a}=2i\alpha\overline{\psi}^{\overline{b}}\Gamma_{\overline{b}a}
\end{equation}
really provide the complex representation of the canonical momenta conjugate to $\psi^{a}$. Let us describe this effectively in terms of real parameters (\ref{eq57a}), (\ref{eq57b}). 

Some remark is necessary here to avoid the conflict (the multiplier $2$) between the last comment concerning (\ref{eq84}) and the formulae (\ref{eq69}) describing the Legendre transformation in terms of complex variables. Namely, (\ref{eq83}) and (\ref{eq84}) describe the effective Darboux representation in the constraints manifold $M=M_{s}$, not in the original non-restricted phase space.

Take the formula (\ref{eq83}) and express it in terms of some fixed basis $(\ldots,e_{a},\ldots)$ introducing the corresponding real coordinates $x^{a}$, $y^{a}$ (\ref{eq57a}). Let us express $\Gamma_{\overline{a}b}$ in terms of this particular basis,
\begin{equation}\label{eq85}
\Gamma_{\overline{a}b}=S_{ab}+iA_{ab},
\end{equation}
where $S_{ab}$, $A_{ab}$ are respectively symmetric and anti-symmetric real matrices,
\begin{equation}\label{eq86}
S_{ab}=S_{ba},\qquad A_{ab}=-A_{ba},
\end{equation}
always related to this particular choice of basis. Then after straightforward calculation one obtains
\begin{equation}\label{eq87}
\gamma||M=\gamma||M_{s}=-2\alpha S_{ab}dx^{a}\wedge dy^{b}-
\alpha A_{ab}\left(dx^{a}\wedge dx^{b}+dy^{a}\wedge dy^{b}\right).
\end{equation}

\noindent{\bf Proposition:} 
If the complex basis $(\ldots,e_{a},\ldots)$ is chosen in such a way that
\begin{equation}\label{eq88}
A_{ab}=0,\qquad S_{ab}=\frac{1}{2\alpha}\delta_{ab}
\end{equation}
(it is always possible when $\Gamma$ is Hermitian and positively definite), then
\begin{equation}\label{eq89}
\gamma||M=\gamma||M_{s}=\delta_{ab}dy^{b}\wedge dx^{a}=dy_{a}\wedge dx^{a}.
\end{equation}

So, we conclude that indeed, from the point of view of real linear structure, $M=M_{s}$ is a real symplectic manifold with Darboux coordinates $x^{a}$ (effective generalized coordinates) and $y_{a}=\delta_{ab}y^{b}$ (generalized conjugate momenta). So indeed, from the point of view of real symplectic geometry, $M=M_{s}$ is a purely second-class manifold (in Dirac language). The effective Hamiltonian responsible for the $L$-dynamics is given on $M=M_{s}$ by
\begin{equation}\label{eq90}
\mathcal{H}=\mathcal{H}_{0}|M=\mathcal{H}_{0}|M_{s},
\end{equation}
i.e., (\ref{eq76}). Let us substitute again the coordinates as above and put
\begin{equation}\label{eq91}
\chi_{\overline{a}b}=\sigma_{ab}+i\alpha_{ab},
\end{equation}
where again $\sigma$, $\alpha$ are real and respectively symmetric and anti-symmetric matrices. We obtain
\begin{equation}\label{eq92}
\mathcal{H}=\frac{\gamma}{2}\sigma_{ab}\left(y^{a}y^{b}+x^{a}x^{b}\right)+
\frac{\gamma}{2}\alpha_{ab}\left(x^{b}y^{a}-x^{a}y^{b}\right).
\end{equation}
This is the real form of the reduced Hamiltonian of dynamics derived from the Schr\"odinger Lagrangian. The equations of motion have the form of the Hamilton equations with the symplectic form (\ref{eq89}), $x^{a}$, $y_{a}=\delta_{ab}y^{b}$ being respectively canonical coordinates and their conjugate momenta, with the standard Poisson brackets.

\noindent{\bf Remark:} 
The first term in (\ref{eq92}) is the usual isotropic harmonic oscillator (isotropic in the sense of $\sigma_{ab}$, not $S_{ab}$, unless both are proportional to each other); the second one is not particularly important because for the Hermitian bilinear forms $\chi_{\overline{a}b}$, $\Gamma_{\overline{a}b}$, when in addition $\Gamma$ is positively definite, it is always possible to choose a basis $(\ldots,e_{a},\ldots)$ in which simultaneously $\Gamma$ is $\delta$-like and $\chi$ is symmetric, i.e., 
\begin{equation}\label{eq92a}
\alpha_{ab}=0, 
\end{equation}
moreover, $\chi$ is then real-diagonal. Formally it is not necessary to assume that $\left[\chi_{\overline{a}b}\right]$, $\left[\sigma_{ab}\right]$ must be positively definite. If they are not, the mentioned interpretation in terms of the usual harmonic oscillator is not literally true.

It is perhaps instructive to express everything in terms of the real variables (\ref{eq57a}), (\ref{eq57b}), although the complex ones (\ref{eq55}), (\ref{eq56}) are formally more convenient, even in field-theoretic problems. We did this partially in formulae (\ref{eq61})--(\ref{eq66}), etc. Let us now look what is the real representation of the Legendre transformation (\ref{eq69}) and some resulting relationships. It is easy to show that the Legendre transformations become
\begin{equation}\label{eq93}
u_{a}=\alpha A_{ab}x^{b}+\alpha S_{ab}y^{b},\qquad 
v_{a}=-\alpha S_{ab}x^{b}+\alpha A_{ab}y^{b}.
\end{equation}

\noindent{\bf Proposition:} 
If the basis $(\ldots,e_{a},\ldots)$ is chosen in such a way that (\ref{eq88}) holds, then one obtains simply that
\begin{equation}\label{eq94}
u_{a}=\frac{1}{2}\delta_{ab}y^{b},\qquad 
v_{a}=-\frac{1}{2}\delta_{ab}x^{b}.
\end{equation}

\noindent{\bf Proposition:} 
It is perhaps a little pretentious and artificially sophisticated, but nevertheless instructive and aesthetic to admit more general bases $(\ldots,e_{a},\ldots)$ in which
\begin{equation}\label{eq95}
S_{ab}=\frac{1}{2\alpha}g_{ab},
\end{equation}
where $g_{ab}=g_{ba}$ and in "physical" models the matrix $\left[g_{ab}\right]$ is positively definite. It plays the role of the Euclidean metric tensor in the $n$-dimensional real linear space $U$ composed of linear combinations $\lambda^{a}e_{a}$ with real coefficients $\lambda^{a}\in\mathbb{R}$ (the $\mathbb{R}$-linear shell of the system of vectors $e_{a}$, $a=\overline{1,n}$). Obviously, the linear quantities
\begin{equation}\label{eq96}
y_{a}:=g_{ab}y^{b}
\end{equation}
may be interpreted as components of $U$-covariant vectors, $y\in U^{\ast}$. Then (\ref{eq87}) becomes 
\begin{equation}\label{eq97}
\gamma||M=\gamma||M_{s}=dy_{a}\wedge dx^{a}-\alpha A_{ab}dx^{a}\wedge dx^{b}-
\alpha \left({}^{g}A\right)^{ab}dy_{a}\wedge dy_{b},
\end{equation}
where ${}^{g}A$ is obtained from $A$ by the $g$-raising of indices,
\begin{equation}\label{eq98}
\left({}^{g}A\right)^{ab}=g^{ac}g^{bd}A_{cd}.
\end{equation}

We avoid to write simply $\left[A^{ab}\right]$ because this might be confused with the contravariant inverse of $\left[A_{ab}\right]$, non-existing in our typically physical situations. In such situations we have just the following form of (\ref{eq89}):
\begin{equation}\label{eq99}
\gamma||M=\gamma||M_{s}=dy_{a}\wedge dx^{a}.
\end{equation}
And (\ref{eq92}) becomes then
\begin{equation}\label{eq100}
\mathcal{H}=\frac{\gamma}{2}\left({}^{g}\sigma\right)^{ab}y_{a}y_{b}+
\frac{\gamma}{2}\sigma_{ab}x^{a}x^{b}+\frac{\gamma}{2}
\left[\left({}^{g}\alpha\right)^{b}{}_{a}-
\left({}^{g}\alpha\right)_{a}{}^{b}\right]x^{a}y_{b},
\end{equation}
where the label "$g$" refers to the $g$-raising of indices,
\begin{equation}\label{eq101}
\left({}^{g}\sigma\right)^{ab}:=g^{ac}\sigma_{cd}g^{db},\qquad
\left({}^{g}\alpha\right)^{b}{}_{a}:=g^{bc}\alpha_{ca},\qquad
\left({}^{g}\alpha\right)_{a}{}^{b}:=\alpha_{ac}g^{cb}.
\end{equation}
The upper-case indices at $g$ refer to the "contravariant inverse", $g^{ac}g_{cb}=\delta^{a}{}_{b}$. We write $\left({}^{g}\sigma\right)^{ab}$ instead of $\sigma^{ab}$ because the latter might be confused with the "contravariant inverse" of $\sigma^{ab}$ which is obviously something else than the $g$-shifted object appearing in (\ref{eq101}). 

\noindent{\bf Remark:} 
The first term in (\ref{eq100}) refers to the kinetic energy of the $x$-oscillator, the second one is its potential energy. The third term is more peculiar. It corresponds to something which formally looks like the constant magnetic field of induction tensor proportional to
\begin{equation}\label{eq101a}
f_{ab}=\gamma\alpha_{ab}=-f_{ba}
\end{equation}
and the covector potential proportional to the linear field
\begin{equation}\label{eq101b}
a_{k}=\gamma\alpha_{kl}x^{l}.
\end{equation}

In this language the Legendre transformation (\ref{eq93}) has the following form:
\begin{equation}\label{eq103}
u_{a}=\alpha A_{ab}x^{b}+\frac{1}{2}g_{ab}y^{b},\qquad v_{a}=-\frac{1}{2}g_{ab}x^{b}+\alpha A_{ab}y^{b},
\end{equation}
and obviously (\ref{eq94}) ($A$ is eliminated) becomes
\begin{equation}\label{eq104}
u_{a}=\frac{1}{2}g_{ab}y^{b},\qquad v_{a}=-\frac{1}{2}g_{ab}x^{b}.
\end{equation}

\subsubsection{Canonical formalism with "direct" nonlinearity}

Quite independently on our quantum motivation (sometimes more or less hidden), the above study is an interesting and instructive example of how the Dirac procedure of Lagrangian constraints works in a rather non-typical situation. One might object here against using the heavy formalism for the geometric discussion of something that is technically simple: non-dissipative linear finite-dimensional object, formally equivalent to some system of coupled harmonic oscillators, perhaps parametrically excited when $\chi$ does depend explicitly on time. It is well known from the theory of linear differential equations in $\mathbb{R}^{k}$-spaces that technically everything is reducible to matrix exponents. Nevertheless, the structures revealed and discussed here may be very useful in the infinite-dimensional case. And besides, they may be a good starting point towards discussing generalized models admitting nonlinearities and second-order time derivatives. Nonlinearities introduced by the $V$-terms in (\ref{eq46}), in particular the ones of the form (\ref{eq49}), (\ref{eq50}), do not modify the above canonical formalism in an essential way. Namely, the Legendre transformation is given by the same formulae (\ref{eq69}) resulting in the same primary constraints $M$ (\ref{eq70}), (\ref{eq71}). According to the formula (\ref{eq72}), the expression for the "energy" (\ref{eq73}) is modified by the $V$-term:
\begin{equation}\label{eq105}
e=\gamma\chi_{\overline{a}b}\overline{\psi}^{\overline{a}}\psi^{b}+
V\left(\psi,\overline{\psi}\right).
\end{equation}
Therefore, the background Hamiltonian (\ref{eq76}) in (\ref{eq75}) is replaced by
\begin{equation}\label{eq106}
\mathcal{H}_{0}=\gamma\chi_{\overline{a}b}\overline{\psi}^{\overline{a}}\psi^{b}+
V\left(\psi,\overline{\psi}\right).
\end{equation}
This implies that the vector field (\ref{eq78}) is modified by the following additive correction term:
\begin{equation}\label{eq107}
X_{V}=-\frac{\partial V}{\partial \psi^{a}}\frac{\partial}{\partial \pi_{a}}-
\frac{\partial V}{\partial \overline{\psi}^{\overline{a}}}\frac{\partial}{\partial \overline{\pi}_{\overline{a}}}.
\end{equation}
For example, for the quartic model
\begin{equation}\label{eq108}
V=A\left(\Gamma_{\overline{a}b}\overline{\psi}^{\overline{a}}\psi^{b}\right)^{2},
\end{equation}
where $A\in\mathbb{R}$ is some constant, we have 
\begin{equation}\label{eq109}
\frac{\partial V}{\partial \psi^{a}}=
2A\left(\overline{\psi}^{\overline{c}}\Gamma_{\overline{c}d}\psi^{d}\right)
\overline{\psi}^{\overline{b}}\Gamma_{\overline{b}a},\qquad
\frac{\partial V}{\partial \overline{\psi}^{\overline{a}}}=
2A\left(\overline{\psi}^{\overline{c}}\Gamma_{\overline{c}d}\psi^{d}\right)
\Gamma_{\overline{a}b}\psi^{b}.
\end{equation}

In virtue of the assumed non-singularity of $\Gamma$, the corresponding equations (\ref{eq75}), (\ref{eq80}) are again uniquely solvable with respect to the Lagrange multipliers at any point of the primary constraints $M$. The only difference in comparison with (\ref{eq81}) is that some additional $V$-dependent terms appear:
\begin{equation}\label{eq110}
\lambda^{a}=-\frac{i}{2}\frac{\gamma}{\alpha}\Gamma^{a\overline{c}}
\chi_{\overline{c}b}\psi^{b}-\frac{i}{2\alpha}\Gamma^{a\overline{c}}
\frac{\partial V}{\partial \overline{\psi}^{\overline{c}}},\quad
\overline{\lambda}^{\overline{a}}=\frac{i}{2}\frac{\gamma}{\alpha}
\overline{\psi}^{\overline{b}}\chi_{\overline{b}c}\Gamma^{c\overline{a}}
+\frac{i}{2\alpha}\frac{\partial V}{\partial \psi^{c}}\Gamma^{c\overline{a}}.
\end{equation}
The existence of these unique solutions all over $M$ implies that again the secondary constraints $M_{s}$ are identical with the primary ones $M$, and that they are second-class constraints, i.e., $M=M_{s}$ is a symplectic manifold in the sense of the structure induced by the symplectic form of the original phase space. 

Referring to formulae (\ref{eq32}) we can simply rewrite (\ref{eq110}) as follows:
\begin{equation}\label{eq111}
\lambda^{a}=-\frac{i}{2}\frac{\gamma}{\alpha}H^{a}{}_{b}\psi^{b}-
\frac{i}{2\alpha}\frac{\partial V}{\partial \psi^{a}},\qquad
\overline{\lambda}^{\overline{a}}=\frac{i}{2}\frac{\gamma}{\alpha}
\overline{\psi}^{\overline{b}}H_{\overline{b}}{}^{\overline{a}}
+\frac{i}{2\alpha}\frac{\partial V}{\partial \overline{\psi}^{\overline{a}}},
\end{equation}
with the suggestive, although not very correct, abbreviation:
\begin{equation}\label{eq112}
\frac{\partial V}{\partial \psi_{a}}:=\Gamma^{a\overline{c}}
\frac{\partial V}{\partial \overline{\psi}^{\overline{c}}},\qquad
\frac{\partial V}{\partial \overline{\psi}_{\overline{a}}}:=\frac{\partial V}{\partial \psi^{c}}\Gamma^{c\overline{a}}.
\end{equation}

\noindent{\bf Remark:} 
Let us be careful: as it is mentioned many times, the "Hamiltonian" (\ref{eq106}) is something completely else than the "true" quantum-mechanical Hamiltonian appearing in the Schr\"odinger equation, even if the non-quadratic term $V$ responsible for the direct nonlinearity does not occur at all. Nevertheless, it is still responsible for the directly nonlinear quantum evolution of $\psi$ as derived from our Lagrangian model.

Using the $M$-reduced canonical momenta (\ref{eq84})
\begin{equation}\label{eq113}
\Pi_{a}=2i\alpha \overline{\psi}^{\overline{b}}\Gamma_{\overline{b}a},\qquad
\overline{\Pi}_{\overline{a}}=-2i\alpha \Gamma_{\overline{a}b}\psi^{b}
\end{equation}
conjugate respectively to $\psi^{a}$ and $\overline{\psi}^{\overline{a}}$ and then eliminating the redundancy, we find again that according to (\ref{eq83}) $\psi$ and $\overline{\psi}$ are respectively (of course up to normalisation) the generalized complex coordinates and their conjugate momenta on $M=M_{s}$, namely 
\begin{equation}\label{eq114}
\left\{\psi^{a},\psi^{b}\right\}_{M}=0,\qquad
\left\{\overline{\psi}^{\overline{a}},\overline{\psi}^{\overline{b}}\right\}_{M}=
0,\qquad \left\{\psi^{a},\overline{\psi}^{\overline{b}}\right\}_{M}=
\frac{1}{2i\alpha}\Gamma^{a\overline{b}}.
\end{equation}
Writing the Hamilton equations of motion on $M=M_{s}$ in the following form:
\begin{equation}\label{eq115}
\frac{d\psi^{a}}{dt}=\left\{\psi^{a},\mathcal{H}\right\}_{M}
\end{equation}
or, equivalently,
\begin{equation}\label{eq116}
\frac{d\overline{\psi}^{\overline{a}}}{dt}=
\left\{\overline{\psi}^{\overline{a}},\mathcal{H}\right\}_{M}
\end{equation}
with the effective Hamiltonian $\mathcal{H}$ on $M=M_{s}$ given by (\ref{eq73}), (\ref{eq74}), (\ref{eq76}) and making use of all standard properties of Poisson brackets, we obtain respectively
\begin{eqnarray}
i\hbar\frac{d\psi^{a}}{dt}&=&H^{a}{}_{b}\psi^{b}+
\frac{1}{2}\Gamma^{a\overline{b}}\frac{\partial V}{\partial \overline{\psi}^{\overline{b}}},
\label{eq117}\\
-i\hbar\frac{d\overline{\psi}^{\overline{a}}}{dt}&=&
\overline{\psi}^{\overline{b}}H_{\overline{b}}{}^{\overline{a}}+
\frac{1}{2}\frac{\partial V}{\partial \psi^{b}}\Gamma^{b\overline{a}},
\label{eq118}
\end{eqnarray}
which are evidently complex conjugates of each other. Obviously, just as previously, we have put
\begin{equation}\label{eq119}
\alpha=\hbar,\qquad \gamma=2,
\end{equation}
even if our interpretation is purely classical. Those are finite-level Schr\"odinger equations with possibly nonlinear terms controlled by $V\left(\psi,\overline{\psi}\right)$. We use the term "direct nonlinearity" to stress the fact that $\Gamma$ is fixed and the possible nonlinearity is just introduced as the perturbation term built of $V$.

This is the model suggested by analytical mechanics in the $W$-space endowed with $\Gamma$-geometry of the Hermitian type. Before we go any further towards some non-direct, geometry-based nonlinearity of non-perturbative type, some comments are necessary.

\subsubsection{Removing Dirac constraints by second-order terms}

Traditional analytical mechanics of classical oscillatory systems with a linear background suggests Lagrangians of the form (\ref{eq38}), quadratic in $\psi$ and perhaps corrected by some anharmonic perturbation (the quartic model of the correction to the Lagrangian seems to be the most popular one, although, of course, some trigonometric, hyperbolic and logarithmic corrections are also used and well motivated). On the other side, the Schr\"odinger-type Lagrangian (\ref{eq26}) describes the usual linear quantum mechanics. Lagrangian (\ref{eq38}) quadratic in derivatives does not imply any Dirac singularity; the Legendre transformation is invertible, assuming of course (what we do) that $\Gamma$ is not degenerate (in standard theory it is just positively definite). Therefore, the natural temptation appears to admit Lagrangians $L(1,2)$ (\ref{eq35}) perhaps with a possible anharmonic correction $V$, when it becomes $L(1,2,{\rm dnl})$ (\ref{eq47}), i.e., the second-order polynomial of velocities with a direct nonlinearity. From the Dirac point of view the term quadratic in velocities introduces some kind of "regularization" because Legendre transformation is then invertible. It is invertible for any, even very small, non-vanishing value of $|\beta|$. Without the $V$-term one obtains from the variation of $\overline{\psi}$ the second-order Schr\"odinger equation (\ref{eq41}); the term "Schr\"odinger" becomes literally true when we put (\ref{eq119}), i.e., $\alpha=\hbar$, $\gamma=2$. The variation with respect to $\psi$ itself leads to the complex-conjugate equation for $\overline{\psi}$. In a sense $\beta$ is an additional "Planck constant". Admitting a direct nonlinearity via the "potential" $V$ in (\ref{eq47}), we obtain the second-order nonlinear "Schr\"odinger equation"
\begin{equation}\label{eq120}
i\hbar\frac{d\psi^{a}}{dt}-\frac{\beta}{2}\frac{d^{2}\psi^{a}}{dt^{2}}=
H^{a}{}_{b}\psi^{b}+\frac{1}{2}\Gamma^{a\overline{b}}\frac{\partial V}{\partial \overline{\psi}^{\overline{b}}}
\end{equation}
and obviously its complex conjugate
\begin{equation}\label{eq121}
-i\hbar\frac{d\overline{\psi}^{\overline{a}}}{dt}-
\frac{\beta}{2}\frac{d^{2}\overline{\psi}^{\overline{a}}}{dt^{2}}=
\overline{\psi}^{\overline{b}}H_{\overline{b}}{}^{\overline{a}}+
\frac{1}{2}\frac{\partial V}{\partial \psi^{b}}\Gamma^{b\overline{a}}.
\end{equation}
In other words, we have the following expression for the variational derivative:
\begin{equation}\label{eq122}
\frac{\delta I(1,2,{\rm dnl})}{\delta\overline{\psi}^{\overline{a}}(t)}=
2i\hbar\Gamma_{\overline{a}b}\frac{d\psi^{b}}{dt}-
\beta\Gamma_{\overline{a}b}\frac{d^{2}\psi^{b}}{dt^{2}}-
\gamma H_{\overline{a}b}\psi^{b}-
\frac{\partial V}{\partial \overline{\psi}^{\overline{a}}},
\end{equation}
where $\gamma=2$ if for $\beta=0$ we are to obtain the usual Schr\"odinger equation. Obviously, performing the variation with respect to $\psi^{a}$, we obtain the complex-conjugate expression
\begin{equation}\label{eq123}
\frac{\delta I(1,2,{\rm dnl})}{\delta\psi^{a}(t)}=
-2i\hbar\frac{d\overline{\psi}^{\overline{b}}}{dt}\Gamma_{\overline{b}a}-
\beta\frac{d^{2}\overline{\psi}^{\overline{b}}}{dt^{2}}\Gamma_{\overline{b}a}-
\gamma\overline{\psi}^{\overline{b}}H_{\overline{b}a}-
\frac{\partial V}{\partial \psi^{a}}.
\end{equation}

\subsubsection{Some physical interpretation of second derivatives}

It is interesting to mention that there is some motivation from other physical problems to admit the term with $\psi$-accelerations, i.e., with the second-order time derivatives introduced to the Schr\"odinger equation. For instance, similar ideas were proposed and studied quite a long time ago by A. Barut and more recently have been re-investigated by V. V. Dvoeglazov, S. Kruglov, J. P. Vigier and others (see, e.g., \cite{Dvoe_05,Krug_04} and references therein). Among others there is also an interesting article where the authors used the analogy between the Schr\"{o}dinger and Fourier equations for nanoscience \cite{JMK_01,JMK_03}. The short description of the idea is presented below.

Hence, the quantum Fourier equation which describes the heat (mass) diffusion on the atomic level has the following form:
\begin{equation}\label{eq+52}
\frac{\partial T}{\partial t}=\frac{\hbar}{m}\nabla^{2}T.
\end{equation}
If we make the substitutions $t\rightarrow it/2$ and $T\rightarrow \psi$, then we end up with the free Schr\"{o}dinger equation:
\begin{equation}\label{eq+53}
i\hbar\frac{\partial \psi}{\partial t}=-\frac{\hbar^{2}}{2m}\nabla^{2}\psi.
\end{equation}
The complete Schr\"{o}dinger equation with the potential term $V$ after the reverse
substitutions $t\rightarrow -2it$ and $\psi\rightarrow T$ gives us the parabolic quantum
Fokker-Planck equation, which describes the quantum heat transport for $\triangle t>\tau$, where $\tau=\hbar/m\alpha^{2}c^{2}\sim 10^{-17}$ sec and $c\tau\sim 1$ nm, i.e.,
\begin{equation}\label{eq+54}
\frac{\partial T}{\partial t}=\frac{\hbar}{m}\nabla^{2}T-\frac{2V}{\hbar}T.
\end{equation}
For ultrashort time processes when $\triangle t<\tau$ one obtains the generalized quantum
hyperbolic heat transport equation
\begin{equation}\label{eq+55}
\tau\frac{\partial^{2} T}{\partial t^{2}}+\frac{\partial T}{\partial
t}=\frac{\hbar}{m}\nabla^{2}T-\frac{2V}{\hbar}T
\end{equation}
(its structure and solutions for ultrashort thermal processes were investigated in \cite{JMK_01}) which leads us to the following second-order modified Schr\"{o}dinger equation:
\begin{equation}\label{eq+56}
i\hbar\frac{\partial \psi}{\partial
t}+2\tau\hbar\frac{\partial^{2}\psi}{\partial t^{2}}=-\frac{\hbar^{2}}{2m}\nabla^{2}\psi+V\psi,
\end{equation}
where the additional term with the second-order time derivative describes the interaction of electrons with surrounding space-time filled with virtual positron-electron pairs. It is easy to see that (\ref{eq+56}) is
analogous to (\ref{eq41}) if we suppose that
\begin{equation}\label{eq+57}
\alpha=\hbar,\qquad \beta=-4\tau\hbar,\qquad \gamma=2.
\end{equation}
{\bf Remark:} obviously, an important question appears: is for such a second-order Schr\"o\-dinger equation something like the probabilistic interpretation still possible? And if not, what is it to be replaced by? Those are still open questions, although some comments will be given below. In a sense the problem is like one of the quantum-mechanical interpretation of the Klein-Gordon equation versus the Schr\"odinger or Dirac equations. In our opinion it is impossible to answer all questions at once, so it is a reasonable way to discuss consequently any geometrically interesting ideas without waiting for their immediate, perhaps premature, physical interpretation. Otherwise one repeats some mistake made by Schr\"odinger who rejected his second-order partial differential equation known today as the Klein-Gordon equation. 

\subsubsection{Regular Legendre transformation and canonical formalism}

If in (\ref{eq35}), (\ref{eq47}) one admits the term quadratic in velocities, i.e., non-vanishing $\beta$, then the Legendre transformation becomes as follows:
\begin{eqnarray}
\pi_{a}&=&\frac{\partial L(1,2,{\rm dnl})}{\partial \dot{\psi}^{a}}=
\frac{\partial L(1,2)}{\partial \dot{\psi}^{a}}=
i\alpha\overline{\psi}^{\overline{b}}\Gamma_{\overline{b}a}+
\beta\dot{\overline{\psi}}^{\overline{b}}\Gamma_{\overline{b}a},
\label{eq124}\\
\overline{\pi}_{\overline{a}}&=&\frac{\partial L(1,2,{\rm dnl})}{\partial \dot{\overline{\psi}}^{\overline{a}}}=
\frac{\partial L(1,2)}{\partial \dot{\overline{\psi}}^{\overline{a}}}=
-i\alpha\Gamma_{\overline{a}b}\psi^{b}+
\beta\Gamma_{\overline{a}b}\dot{\psi}^{b}.
\label{eq125}
\end{eqnarray}
The corresponding "energy" function of Lagrangian model is given by 
\begin{equation}\label{eq126}
e=\beta\Gamma_{\overline{a}b}\dot{\overline{\psi}}^{\overline{a}}\dot{\psi}^{b}
+\gamma\chi_{\overline{a}b}\overline{\psi}^{\overline{a}}\psi^{b}+
V\left(\psi,\overline{\psi}\right).
\end{equation}
No constraints in the classical phase space appear, so the Legendre transformation is invertible and its inverse has the following form:
\begin{equation}\label{eq127-128}
\dot{\psi}^{a}=\frac{1}{\beta}\Gamma^{a\overline{b}}\overline{\pi}_{\overline{b}}
+\frac{i\alpha}{\beta}\psi^{a},
\qquad
\dot{\overline{\psi}}^{\overline{a}}=
\frac{1}{\beta}\pi_{b}\Gamma^{b\overline{a}}
-\frac{i\alpha}{\beta}\overline{\psi}^{\overline{a}}.
\end{equation}
The corresponding "Hamiltonian" is globally defined on the phase space:
\begin{equation}\label{eq129}
\mathcal{H}=\frac{1}{\beta}\left(\Gamma^{a\overline{b}}\pi_{a}
\overline{\pi}_{\overline{b}}+i\alpha\left[\pi_{a}\psi^{a}-
\overline{\pi}_{\overline{a}}\overline{\psi}^{\overline{a}}\right]\right)+
\left(\frac{\alpha^{2}}{\beta}\Gamma_{\overline{a}b}+
\gamma\chi_{\overline{a}b}\right)\overline{\psi}^{\overline{a}}\psi^{b}+
V\left(\psi,\overline{\psi}\right).
\end{equation}
For any $\beta\neq 0$, this is a regular Hamiltonian and equations of motion may be written in the usual canonical form:
\begin{equation}\label{eq130}
\frac{d\psi^{a}}{dt}=\left\{\psi^{a},\mathcal{H}\right\}=\frac{\partial \mathcal{H}}{\partial \pi_{a}},\qquad \frac{d\pi_{a}}{dt}=\left\{\pi_{a},\mathcal{H}\right\}=-\frac{\partial \mathcal{H}}{\partial \psi^{a}},
\end{equation}
or in the equivalent form of complex-conjugate quantities:
\begin{equation}\label{eq131}
\frac{d\overline{\psi}^{\overline{a}}}{dt}=
\left\{\overline{\psi}^{\overline{a}},\mathcal{H}\right\}=\frac{\partial \mathcal{H}}{\partial \overline{\pi}_{\overline{a}}},\qquad \frac{d\overline{\pi}_{\overline{a}}}{dt}=
\left\{\overline{\pi}_{\overline{a}},\mathcal{H}\right\}=-\frac{\partial \mathcal{H}}{\partial \overline{\psi}^{\overline{a}}}.
\end{equation}
There are no phase-space constraints and no Lagrange multipliers.

The limit transition $\beta\rightarrow 0$ is a rather complicated and obscure problem. The equations of motion (\ref{eq120}), (\ref{eq121}) transform then smoothly into (\ref{eq117}), (\ref{eq118}). Obviously, this concerns the very form of equations, but the phase portraits change catastrophically as it is usually the case when the highest-order derivative terms in equations are neglected. The energy expression (\ref{eq126}) also reduces smoothly to (\ref{eq105}). But it is no longer the case with the Hamiltonian (\ref{eq129}). There is no directly well-defined limit when $\beta$ in (\ref{eq129}) tends to zero. In any case nothing like (\ref{eq75}), (\ref{eq76}), (\ref{eq106}) appears. This is due to the fact that for $\beta=0$ the Legendre transformation is not invertible. For $\beta\neq 0$ the number of complex degrees of freedom remains $n$, i.e., that of real ones remains $2n$. But for the vanishing $\beta$, the number of effective real degrees of freedom becomes $n$. This qualitative discontinuity resembles similar phenomena in field theory, e.g., passing over from the non-vanishing mass of "photons" in the Proca equations to the vanishing one in the Maxwell ones.

In this sense the non-vanishing $\beta$ leads to some "regularization" of the underlying Hamilton mechanics, but it is not yet clear how this regularization interferes with the probabilistic interpretation, the structure of the effective scalar product of "wave functions" and its positive definiteness. This resembles the corresponding problem for the Klein-Gordon equation.

\section{Non-direct nonlinearity and Hamiltonian systems on manifolds of se\-cond-order tensors}

The mentioned above models of nonlinearity in a finite-level Schr\"odinger equation or second-order "Schr\"odinger equation" were "direct" in the sense that the term of Lagrangian responsible for nonlinear phenomena was a kind of perturbation extra introduced as an additive correction to the linear background. There is a rather large freedom of a priori admissible models and no very convincing criteria of choice do exist. The choice is indeed a matter of intuition and is based on a kind of phenomenology. And because of this it is never very convincing.

Below we try to discuss some models where nonlinearity is not introduced "by hand", but rather it is based on some kind of geometric aprioric arguments.

\subsection{Removing absolute objects and admitting dynamical scalar product}

Qualitatively, the idea is as follows: Lagrangians (\ref{eq26}), (\ref{eq35}), (\ref{eq36}), (\ref{eq46}), (\ref{eq47}) contain only one dynamical quantity $\psi$ with $n$ complex degrees of freedom, i.e., $2n$ real ones $(x,y)$. As we saw, it is convenient to follow the method popularly used in field theory and consider formally the components $\psi^{a}$ and their $\overline{\psi}^{\overline{a}}$ as independent variables. But all the mentioned Lagrangians contain also some absolute (non-dynamical) object, namely, the scalar product $\Gamma$. It is fixed once for all, just like the metric tensor of Euclidean space or that of Minkowskian space-time of special relativity. And it is just here some doubts appear. Namely, 
\begin{center}
{\bf Nature does not like absolute objects.} 
\end{center}
It turned out that in general relativity the metric tensor of our four-dimensional (in general curved) space-time is a dynamical quantity which together with "physical" or "matter" fields satisfies a closed system of differential equations. This system is based on the mutual interaction and the metric tensor turns out to be also some kind of a physical field, namely, it just describes relativistic gravitation. Hence, there are no absolute objects in space-time. 
\medskip

{\em One can ask why not to follow this pattern and admit $\Gamma$ to be a dynamical object mutually interacting with $\psi$. For what would be the physical reality hidden behind some absolutely fixed $\Gamma$?}
\medskip

It turns out that the only natural Lagrangian dynamics of the $\Gamma$-object is strongly nonlinear and this introduces a kind of effective nonlinearity to the dynamics of the total $\left(\Gamma,\psi\right)$-system. This nonlinearity has geometric origin, therefore, it is well motivated and non-perturbative. By that we mean that it is not a correction to any well-defined linear background and because of this it cannot be analysed by standard techniques of the perturbation calculus, any expansion with respect to some "small" parameter, etc.

Analytically $\Gamma$ is represented by the quadratic matrix with coefficients $\Gamma_{\overline{a}b}$. Geometrically it is a Hermitian element of the tensor space $\overline{W}^{\ast}\otimes W^{\ast}$, i.e., a Hermitian form on $W$. In various problems of analytical mechanics and mathematical physics one often deals with dynamical systems the phase spaces of which are byproducts of some matrix manifolds. Obviously, those matrices appear on the level of calculus and analytical representation but as a matter of fact they represent various second-order tensors, i.e., linear mappings operating between two linear spaces. In applications which present interest for us they are usually isomorphisms between linear spaces of the same dimension, i.e., we deal with quadratic non-singular matrices. Some exceptional models when one deals with rectangular (not necessary quadratic) matrices are presented in \cite{Roz_05}.

\subsubsection{Manifolds of linear mappings}

Let us begin with some introductory remarks. Matrices provide analytical description of linear mappings. So, let $U$, $V$ be some linear spaces of the same dimension. At a moment, we do not precise if they are meant over the real of complex field. Let ${\rm L}(U,V)$ denote the linear space of all linear mappings of $U$ into $V$ and let ${\rm LI}(U,V)\subset{\rm L}(U,V)$ denote its open submanifold consisting of linear isomorphisms of $U$ onto $V$. When dealing with analytical formulae, we shall use some bases $\left(\ldots,E_{A},\ldots\right)$, $\left(\ldots,e_{i},\ldots\right)$ respectively in $U$ and $V$. Linear mappings $\varphi\in {\rm L}(U,V)$ are analytically represented by matrices with coefficients $\varphi^{i}{}_{A}$ meant in the following convention:
\begin{equation}\label{eq132}
\varphi E_{A}=e_{i}\varphi^{i}{}_{A}.
\end{equation}
The conjugate mappings $\varphi^{\ast}:V^{\ast}\rightarrow U^{\ast}$ is given by
\begin{equation}\label{eq133}
\varphi^{\ast}p=p\circ \varphi,\qquad \varphi^{\ast}e^{i}=\varphi^{i}{}_{A}E^{A},
\end{equation}
where $\left(\ldots,E^{A},\ldots\right)$, $\left(\ldots,e^{i},\ldots\right)$ denote as usually the dual bases of dual spaces $U^{\ast}$, $V^{\ast}$:
\begin{equation}\label{eq134}
E^{A}\left(E_{B}\right)=\left\langle E^{A},E_{B}\right\rangle=\delta^{A}{}_{B},\qquad e^{i}\left(e_{j}\right)=\left\langle e^{i},e_{j}\right\rangle=\delta^{i}{}_{j}.
\end{equation}
And as usual we make use of abbreviations:
\begin{eqnarray}
&&{\rm L}(U):={\rm L}(U,U), \qquad {\rm L}(V):={\rm L}(V,V),\qquad {\rm etc.} \label{eq135}\\
&&{\rm GL}(U):={\rm LI}(U,U), \qquad {\rm GL}(V):={\rm LI}(V,V),\qquad {\rm etc.} \label{eq136}
\end{eqnarray}
Obviously, ${\rm L}(U)$, ${\rm L}(V)$ are associative algebras and at the same time commutator Lie algebras. They are of course canonically identical with Lie algebras of the corresponding Lie groups ${\rm GL}^{+}(U)$, ${\rm GL}^{+}(V)$, i.e., the connected components of ${\rm GL}(U)$, ${\rm GL}(V)$.

In applications we often deal with various closed submanifolds of ${\rm L}(U,V)$, usually distinguished by some additional structures in $U$, $V$ like, e.g., scalar products. First of all, let us stress that the groups ${\rm GL}(U)$, ${\rm GL}(V)$ act in a natural way on ${\rm LI}(U,V)$, respectively on the right and on the left. So, for any $A\in {\rm GL}(V)$, $B\in {\rm GL}(U)$ we have the action:
\begin{equation}\label{eq137}
{\rm LI}(U,V)\ni\varphi\rightarrow A\varphi B=\left(L_{A}\circ R_{B}\right)(\varphi)=\left(R_{B}\circ L_{A}\right)(\varphi),
\end{equation}
where the symbols $L_{A}$, $R_{B}$ refer respectively to the left- and right-hand-side actions. Obviously, both $L_{{\rm GL}(V)}$, $R_{{\rm GL}(U)}$ act effectively and freely on ${\rm LI}(U,V)$. However, it is important that ${\rm GL}(V)\times {\rm GL}(U)$ does not act effectively through $L_{{\rm GL}(V)}R_{{\rm GL}(U)}$, because dilatations in ${\rm GL}(V)$ and ${\rm GL}(U)$ act in the same way on ${\rm LI}(U,V)$ and the kernel of non-effectiveness of ${\rm GL}(V)\times {\rm GL}(U)$ in the action through (\ref{eq137}) is given by the subgroup 
\begin{equation}\label{eq138}
\left\{\left(\lambda{\rm Id}_{V},\lambda^{-1}{\rm Id}_{U}
\right):\lambda\in\mathbb{R}\backslash \{0\}\right\}.
\end{equation}
Let us mention that ${\rm L}(U)$, ${\rm L}(V)$ are also semigroups under the composition of mappings and these semigroups act on the total ${\rm L}(U,V)$ according to the formula (\ref{eq137}). Obviously, if $A$, $B$ are not isomorphisms, i.e., elements of ${\rm GL}(V)$, ${\rm GL}(U)$, the submanifold ${\rm LI}(U,V)$ is not preserved by them. If linear spaces $U$, $V$ are endowed with some scalar products, i.e., symmetric (usually non-degenerate) bilinear forms, $\eta\in U^{\ast}\otimes U^{\ast}$, $g\in V^{\ast}\otimes V^{\ast}$, then the special attention is paid to submanifolds of "isometries" ${\rm O}(U,\eta;V,g)\subset{\rm LI}(U,V)$, consisting of such mappings $\varphi\in{\rm LI}(U,V)$ that
\begin{equation}\label{eq139}
g=\varphi^{\ast}\eta,\qquad g_{ij}=\eta_{AB}\varphi^{A}{}_{i}\varphi^{B}{}_{j}.
\end{equation}
They are homogeneous spaces of isometry groups ${\rm O}(V,g)$, ${\rm O}(U,\eta)$ acting in the sense of (\ref{eq137}). These isometry groups consist of transformations preserving respectively $g$ and $\eta$,
\begin{equation}\label{eq140}
g=A^{\ast}g,\qquad \eta=B^{\ast}\eta,
\end{equation}
i.e., analytically,
\begin{equation}\label{eq141}
g_{ij}=g_{kl}A^{k}{}_{i}A^{l}{}_{j},\qquad \eta_{KL}=\eta_{MN}B^{M}{}_{K}B^{N}{}_{L}.
\end{equation}
In such applications we usually deal with real linear spaces and symmetric positively definite metrics. Nevertheless, everything works well at this stage also for general tensors $\eta$, $g$, even without any kind of symmetry, or, in the extreme case, when $\eta$, $g$ are skew-symmetric, i.e., if one deals with symplectic structures in $U$, $V$. Another interesting case is when one considers as relevant submanifolds of ${\rm LI}(U,V)$ the orbits of the subgroups ${\rm U}(V)\times {\rm U}(U)$, ${\rm SL}(V)\times {\rm SL}(U)$ acting through (\ref{eq137}). Here ${\rm U}(V)$, ${\rm U}(U)$ and ${\rm SL}(V)$, ${\rm SL}(U)$ denote respectively the unimodular and special linear subgroups of ${\rm GL}(V)$, ${\rm GL}(U)$, i.e., groups consisting of linear mappings the determinants of which respectively have modulus one or are just themselves equal to one. If $U$, $V$ are real linear spaces, then ${\rm U}(V)$, ${\rm U}(U)$ preserve the volumes, i.e., the Lebesgue measures in $V$, $U$ (it does not matter how normalised), and ${\rm SL}(V)$, ${\rm SL}(U)$ preserve both the volumes and orientations (both standards of orientation separately). The manifold ${\rm LI}(U,V)$ is then foliated into orbits which are homogeneous spaces of the mentioned subgroup acting through (\ref{eq137}). In the case of real spaces the strata of the action of unimodular groups have two connected components corresponding to the fate of orientation under the group action.

\subsubsection{Hermitian metrics and unitary mappings}

If linear spaces $U$, $V$ are complex, then in applications we are interested rather in sesquilinear Hermitian products $\eta$, $g$ than in bilinear ones. The orthogonal groups are then replaced by the unitary ones ${\rm U}(U,\eta)$, ${\rm U}(V,g)$. They consist of transformations $A\in{\rm U}(V,g)\subset{\rm GL}(V)$, $B\in{\rm U}(U,\eta)\subset{\rm GL}(U)$ satisfying (\ref{eq140}) in the analytical form:
\begin{equation}\label{eq142}
g_{\overline{i}j}=g_{\overline{k}l}\overline{A}^{\overline{k}}{}_{\overline{i}}
A^{l}{}_{j},\qquad \eta_{\overline{K}L}=\eta_{\overline{M}N}
\overline{B}^{\overline{M}}{}_{\overline{K}}B^{N}{}_{L}.
\end{equation}
The Hermitian metrics $g\in \overline{V}^{\ast}\otimes V^{\ast}$, $\eta\in \overline{U}^{\ast}\otimes U^{\ast}$ give rise to the manifold of unitary isometries ${\rm U}(U,\eta;V,g)$; it consists of linear mappings relating $\eta$ to $g$,
\begin{equation}\label{eq143}
\eta=\varphi^{\ast}g,\qquad \eta_{\overline{A}B}=g_{\overline{i}j}
\overline{\varphi}^{\overline{i}}{}_{\overline{A}}\varphi^{j}{}_{B}.
\end{equation}
Obviously, the subset ${\rm U}(U,\eta;V,g)$ is a homogeneous space of both ${\rm U}(V,g)$, ${\rm U}(U,\eta)$ acting through (\ref{eq137}). In applications we are interested in, one deals usually with Hermitian positively definite metrics $\eta$, $g$, but some general statements are valid without this restriction, i.e., pseudo-unitary groups are formally admissible. Similarly, if $U$, $V$ are real linear spaces, then usually $\eta$, $g$ are symmetric positively definite forms. It is so in mechanics of affinely-rigid bodies investigated by us and others in various papers \cite{Bur_08,Cas_95,Chev_04,Kan_04,Pap_01,all_04,all_05,Sol_00,Dias_94} (then $U$, $V$ are translation spaces respectively in the material and physical affine spaces).

\subsection{Non-holonomic velocities and invariants}

Let $Q\subset{\rm LI}(U,V)$ be a submanifold used as a configuration space of some "analytical mechanics". If $U$, $V$ are complex, then $Q$ may (however need not to) be analytic. Moreover, in applications we have in mind, it usually is non-analytic. "Motions" are described by curves 
\begin{equation}\label{eq144}
\mathbb{R}\ni t\rightarrow\varphi(t)\in Q\subset{\rm LI}(U,V).
\end{equation}
Both from the point of view of geometrical foundations and practical calculations, it is often convenient to use non-holonomic velocities which at the time instant $t\in\mathbb{R}$ are given by
\begin{eqnarray}
\Omega(t)&:=&\frac{d\varphi}{dt}(t)\varphi^{-1}(t)\in{\rm L}(V),
\label{eq145}\\
\widehat{\Omega}(t)&:=&\varphi^{-1}(t)\frac{d\varphi}{dt}(t)
=\varphi^{-1}(t)\Omega(t)\varphi(t)\in{\rm L}(U),
\label{eq146}
\end{eqnarray}
i.e., analytically,
\begin{equation}\label{eq147}
\Omega^{i}{}_{j}=\frac{d\varphi^{i}{}_{A}}{dt}\varphi^{-1A}{}_{j},\qquad
\widehat{\Omega}^{A}{}_{B}=\varphi^{-1A}{}_{i}\frac{d\varphi^{i}{}_{B}}{dt}.
\end{equation}
If the manifold $Q\subset{\rm LI}(U,V)$ is parameterized by generalized coordinates $q^{\mu}$, $\mu=1,\ldots,\dim Q$, then obviously
\begin{equation}\label{eq148}
\Omega^{i}{}_{j}\left(q,\dot{q}\right)=
\Omega^{i}{}_{j\mu}(q)\frac{dq^{\mu}}{dt},\qquad
\widehat{\Omega}^{A}{}_{B}\left(q,\dot{q}\right)=
\widehat{\Omega}^{A}{}_{B\mu}(q)\frac{dq^{\mu}}{dt},
\end{equation}
where
\begin{equation}\label{eq149}
\Omega^{i}{}_{j\mu}=\frac{\partial\varphi^{i}{}_{A}}{\partial q^{\mu}}\varphi^{-1A}{}_{j},\qquad
\widehat{\Omega}^{A}{}_{B\mu}=\varphi^{-1A}{}_{i}
\frac{\partial\varphi^{i}{}_{B}}{\partial q^{\mu}}.
\end{equation}
The advantage of using the quantities $\Omega$, $\widehat{\Omega}$ instead of the generalized velocities $\dot{q}^{\mu}$ is that due to their tensorial structure one can construct of them some scalar invariants:
\begin{equation}\label{eq150}
I_{p}:={\rm Tr}\left(\Omega^{p}\right)={\rm Tr}\left(\widehat{\Omega}^{p}\right),\qquad p=1,\ldots,\dim V;
\end{equation}
according to the Cayley-Hamilton theorem, taking other values of $p$ one does not obtain anything new, just some functions of (\ref{eq150}). Obviously, the most important are expressions quadratic in velocities, i.e., $I_{2}$ and $\left(I_{1}\right)^{2}$, because in analytical mechanics they are used for constructing kinetic energy models ($I_{2}$ is the main term and $I_{1}$ is a merely correction). Even for some non-quadratic (in velocities) hypothetical models of the kinetic energy $T$, the quantities $I_{p}$, first of all $I_{2}$ and $\left(I_{1}\right)^{2}$, are reasonable modules for constructing $T$ in a form, e.g.,
\begin{equation}\label{eq151}
T=f\left(I_{2},\left(I_{1}\right)^{2}\right),
\end{equation}
where $f$ is some appropriately postulated function of two variables (incidentally, let us mention that in the three-dimensional spatial formulation of relativistic mechanics of the material point the kinetic Lagrangian is an irrational function of the squared absolute value of velocity).

\subsubsection{Kinetic energies, i.e., Riemannian metrics on manifolds of linear mappings. High-symmetry models}

Quadratic forms of velocities tangent to $Q\subset {\rm LI}(U,V)$ are geometrically equivalent to some metric tensors on $Q$. Usually, at least in our problems, $Q$ is somehow special from the point of view of geometry of ${\rm LI}(U,V)$ and the most natural metrics on $Q$ (viable models of kinetic energy) are restrictions to $Q$ of geometrically distinguished metric tensors on the total ${\rm LI}(U,V)$. And this is where we must begin with some digression concerning our former work on analytical mechanics of affinely-rigid bodies, because some ideas developed there seem to be an inspiration for our search of geometric Schr\"odinger nonlinear models. Let us remind that if $U$, $V$ are real linear spaces of material and spatial translations respectively, then the usual formula for the kinetic energy of internal (relative) degrees of freedom reads \cite{all_04,all_05}
\begin{equation}\label{eq152}
T=\frac{1}{2}g_{ij}\frac{d\varphi^{i}{}_{A}}{dt}\frac{d\varphi^{j}{}_{B}}{dt}
J^{AB},
\end{equation}
where $g\in V^{\ast}\otimes V^{\ast}$ is the metric tensor of the physical space and $J\in U\otimes U$ is, roughly speaking, the constant co-moving tensor of inertia (the second-order tensor moment of the mass distribution with respect to the Lagrange coordinates). This expression is invariant under (\ref{eq137}), where $A$, $B$ are confined respectively to ${\rm O}(V,g)$, ${\rm O}\left(U,\widetilde{J}\right)$. In the special case of inertially isotropic affine body, when $\widetilde{J}$ is proportional to the reference material metric $\widetilde{\eta}$ in the contravariant reciprocal form, thus,
\begin{equation}\label{eq153}
J^{AB}=I\eta^{AB},\qquad \widetilde{J}_{AC}J^{CB}=\delta_{A}{}^{B},
\end{equation}
the expression (\ref{eq152}) is isotropic both spatially and materially. If $\varphi$ is an isometry, i.e., $\varphi\in{\rm O}\left(U,\widetilde{\eta};V,g\right)$, then (\ref{eq152}) becomes the usual kinetic energy of the metrically-rigid body (gyroscope) and the resulting metric tensor is just the restriction of the metric underlying (\ref{eq152}) in ${\rm LI}(U,V)$ to the submanifold of isometries. Just as the latter one, the metric of (\ref{eq152}) is not invariant under the total group (\ref{eq137}). And it is just here where the quasivelocities (\ref{eq145})--(\ref{eq147}) become interesting. Namely, they transform under (\ref{eq137}) according to the following rule:
\begin{equation}\label{eq154}
\Omega\rightarrow A\Omega A^{-1},\qquad \widehat{\Omega}\rightarrow B^{-1}\widehat{\Omega}B.
\end{equation}
Because of this, the corresponding quantities $I_{p}$ (\ref{eq150}) are invariant under the total (\ref{eq137}), i.e., under the most natural group underlying geometry of degrees of freedom. Therefore, the corresponding kinetic Lagrangians built of $I_{p}$, in particular the quadratic ones built of $I_{2}$ and $\left(I_{1}\right)^{2}$, and their underlying metric tensors are also affinely invariant, i.e., non-sensible to the action of (\ref{eq137}). The underlying metric tensors on ${\rm LI}(U,V)$ are essentially Riemannian (they have non-vanishing Riemann tensors). This curvature is due to the fact that in expression for $T$ given by combinations of $I_{2}$ and $\left(I_{1}\right)^{2}$ with constant coefficients, the corresponding quadratic forms of generalized velocities $d\varphi^{i}{}_{A}/dt$ have irreducibly $\varphi$-dependent coefficients. And no change of generalized coordinates for some new ones $q^{\mu}$ may help here. The corresponding quadratic forms of $\dot{q}^{\mu}$ will have always $q$-dependent coefficients. The reason is that $\Omega$, $\widehat{\Omega}$ are essentially non-holonomic quantities and this in turn follows from the non-commutativity of linear groups ${\rm GL}(V)$, ${\rm GL}(U)$.

\noindent{\bf Proposition:} 
As mentioned, from the point of view of geometric a priori, the most natural models of kinetic energies have the following form:
\begin{equation}\label{eq155}
T=\frac{A}{2}{\rm Tr}\left(\Omega^{2}\right)+\frac{B}{2}\left({\rm Tr}\ \Omega\right)^{2}=\frac{A}{2}{\rm Tr}\left(\widehat{\Omega}^{2}\right)+
\frac{B}{2}\left({\rm Tr}\ \widehat{\Omega}\right)^{2}.
\end{equation}
This expression and the underlying metric tensor on ${\rm LI}(U,V)$ are invariant under the total (\ref{eq137}). And this is the most general metric of this property. This is an affine counterpart for the kinetic energy of the spherical metrically-rigid body. 

\noindent{\bf Proposition:} 
Of course, in principle, just like in mentioned rigid body mechanics, one can think about models affinely invariant in space or ones affinely invariant in the material of the body. As in \cite{all_04,all_05} they are respectively given by
\begin{eqnarray}
T&=&\frac{1}{2}\mathcal{L}^{B}{}_{A}{}^{D}{}_{C}
\widehat{\Omega}^{A}{}_{B}\widehat{\Omega}^{C}{}_{D},\label{eq156}\\
T&=&\frac{1}{2}\mathcal{R}^{j}{}_{i}{}^{l}{}_{k}
\Omega^{i}{}_{j}\Omega^{k}{}_{l},\label{eq157}
\end{eqnarray}
obviously with constant coefficients $\mathcal{L}$, $\mathcal{R}$.

\noindent{\bf Remark:} 
The only situation of (\ref{eq156}) and (\ref{eq157}) to coincide, i.e., of the invariance under the total (\ref{eq137}), is just (\ref{eq155}), the geometrically most natural situation. One should stress that (\ref{eq155}) is not positively definite, its main term $\left(A/2\right){\rm Tr} \left(\Omega^{2}\right)$ has the signature $\left(n(n+1)/2\ +,n(n-1)/2\ -\right)$, so, if gyroscopic constraints are taken into account, $A$ must be negative. It was shown in our earlier papers on affine bodies that the non-definiteness of (\ref{eq155}) may be just convenient in certain models of elastic vibrations; in those models the dynamics of deformative motion is encoded in the kinetic energy term, i.e., in some kind of the effective metric on ${\rm LI}(U,V)$. This resembles in a sense the idea of Maupertuis principle. 

Obviously, in certain phenomenological models of deformative dynamics and the rotation-deformation coupling it may be reasonable to try to mix (\ref{eq155}) (or (\ref{eq156}) and (\ref{eq157})) with the metrical term (\ref{eq152}), especially with its isotropic form (\ref{eq153}). 

\noindent{\bf Proposition:}
In certain problems it may be perhaps reasonable to postulate the affine model (\ref{eq155}) perturbed by two corrections breaking the background affine symmetry to the orthogonal one, both in $(V,g)$ and $(U,\eta)$, e.g., \cite{all_04,all_05}
\begin{eqnarray}
T&=&\frac{I_{1}}{2}g_{ik}g^{jl}\Omega^{i}{}_{j}\Omega^{k}{}_{l}+
\frac{I_{2}}{2}\eta_{KL}\eta^{MN}\widehat{\Omega}^{K}{}_{M}
\widehat{\Omega}^{L}{}_{N}+\frac{A}{2}{\rm Tr}\left(\Omega^{2}\right)+
\frac{B}{2}\left({\rm Tr}\ \Omega\right)^{2}\qquad \label{eq158}\\
&=&\frac{I_{1}}{2}g_{ik}g^{jl}\Omega^{i}{}_{j}\Omega^{k}{}_{l}+
\frac{I_{2}}{2}\eta_{KL}\eta^{MN}\widehat{\Omega}^{K}{}_{M}
\widehat{\Omega}^{L}{}_{N}+\frac{A}{2}{\rm Tr}\left(\widehat{\Omega}^{2}\right)+
\frac{B}{2}\left({\rm Tr}\ \widehat{\Omega}\right)^{2}.\nonumber
\end{eqnarray}

If we use orthonormal coordinates in which $g_{ik}=_{\ast}\delta_{ik}$, $\eta_{AB}=_{\ast}\delta_{AB}$, then the following simple formula based on the $\mathbb{R}^{n}$-object may be used for (\ref{eq158}):
\begin{eqnarray}
T&=&\frac{I_{1}}{2}{\rm Tr}\left(\Omega^{T}\Omega\right)+
\frac{I_{2}}{2}{\rm Tr}\left(\widehat{\Omega}^{T}\widehat{\Omega}\right)
+\frac{A}{2}{\rm Tr}\left(\Omega^{2}\right)+
\frac{B}{2}\left({\rm Tr}\ \Omega\right)^{2}\nonumber\\
&=&\frac{I_{1}}{2}{\rm Tr}\left(\Omega^{T}\Omega\right)+
\frac{I_{2}}{2}{\rm Tr}\left(\widehat{\Omega}^{T}\widehat{\Omega}\right)
+\frac{A}{2}{\rm Tr}\left(\widehat{\Omega}^{2}\right)+
\frac{B}{2}\left({\rm Tr}\ \widehat{\Omega}\right)^{2}.\label{eq159}
\end{eqnarray}

\noindent{\bf Remark:} 
Let us stress the following important circumstance. Unlike (\ref{eq158}), (\ref{eq159}), the expression (\ref{eq155}) does not preassume any fixed metrics in $U$, $V$. This will be just the pattern to be followed in our model of "non-direct" geometric nonlinearity in dynamical systems motivated by the Schr\"odinger equation. There we deal, of course, with complex linear spaces. The above formulae, although providing some guiding hints, cannot be literally used because the corresponding Lagrangians would be either trivial or complex. Neither their real nor imaginary parts
\begin{equation}\label{eq160}
{\rm Re}\ T=\frac{1}{2}\left(T+\overline{T}\right),\qquad {\rm Im}\ T=\frac{1}{2i}\left(T-\overline{T}\right)
\end{equation}
would be useful; they do not correspond to any expressions interpretable in quantum-mechanical terms.

As yet, the linear spaces $U$, $V$ were assumed completely unrelated to each other. And it was just correct in the mentioned applications to elastodynamics. Let us remind that ${\rm L}(U)$, ${\rm L}(V)$ are canonically isomorphic with the commutator-sense Lie algebras of ${\rm GL}(U)$, ${\rm GL}(V)$ and the expressions (\ref{eq150}) were their Casimir invariants; in particular, the appropriate special case of (\ref{eq155}), i.e.,
\begin{equation}\label{eq158_2}
A=2n,\qquad B=-2,
\end{equation}
corresponds to the Killing metric (degenerate on the total ${\rm L}(U)$, ${\rm L}(V)$ because those algebras are not semisimple; dilatations form the normal subgroups of ${\rm GL}(U)$, ${\rm GL}(V)$). In the mentioned elastodynamical applications we often deal with the situation where the configuration space $Q\subset {\rm LI}\left(U,V\right)$ is an orbit of some subgroups $G_{U}\subset{\rm GL}(U)$, $G_{V}\subset{\rm GL}(V)$ acting through (\ref{eq137}). Those subgroups are isomorphic and for any $\phi\in Q$ we have that
\begin{equation}\label{eq159_2}
G_{U}=\phi^{-1}G_{V}\phi,\qquad G_{V}=\phi G_{U}\phi^{-1}.
\end{equation}

\noindent{\bf Remark:} 
The usual, i.e., metrically-rigid, body is a typical example, we have then
\begin{equation}\label{eq160_2}
G_{U}={\rm O}\left(U,\eta\right),\qquad G_{V}={\rm O}\left(V,g\right),\qquad \phi\in{\rm O}\left(U,\eta;V,g\right).
\end{equation}
More precisely, in some realistic mechanical applications we are dealing then with the connected components ${\rm O}^{+}\left(U,\eta\right)={\rm SO}\left(U,\eta\right)$, ${\rm O}^{+}\left(V,g\right)={\rm SO}\left(V,g\right)$ and their orbits. Similarly, one considers orbits of ${\rm SL}(U)$, ${\rm SL}(V)$ (incompressible body), etc. Restricting the expressions (\ref{eq150}) to Lie subalgebras $G^{\prime}_{U}\subset{\rm L}(U)$, $G^{\prime}_{V}\subset{\rm L}(V)$ one obtains some Casimir invariants; usually they need not be independent. For example, for ${\rm SO}\left(U,\eta\right)$, ${\rm SO}\left(V,g\right)$ we have $I_{p}=0$ for any odd $p$.

Before going any further we must discuss some special situations. As mentioned, the linear spaces $U$, $V$ above where independent on each other. The corresponding configuration spaces consisted of linear mappings $\varphi\in{\rm L}(U,V)$, i.e., of tensor quantities $\varphi\in V\otimes U^{\ast}$; they were analytically represented by matrices $\varphi^{i}{}_{A}$ (doubled quantities in the Schouten-Veblen language). But when discussing Hamiltonian systems inspired by the Schr\"odinger equation, we must use matrices which analytically represent some scalar products, i.e., twice covariant tensors in $W$, more precisely, the sesquilinear Hermitian forms $\Gamma\in\overline{W}^{\ast}\otimes W^{\ast}$ represented analytically by matrices $\left[\Gamma_{\overline{a}b}\right]$ (in realistic applications positively definite ones). Some relatively new structures appear then.

\subsubsection{Metrics on groups as the special case}

Before discussing the manifolds of scalar products and the aforementioned structures, it is instructive to start, for comparison and for preparing the proper mathematical instruments, with the situation $U=V$. Then ${\rm LI}(U,V)$ becomes simply the linear group ${\rm GL}(V)={\rm GL}(U)$ and the resulting scheme is that of (more or less) invariant Hamiltonian systems on the total Lie groups or their subgroups with some interplay of left and right invariance. Configuration space consists of non-degenerate mixed second-order tensors (once contravariant and once covariant) in a given linear space. The corresponding metric tensors underlying the kinetic energy expressions like (\ref{eq156}), (\ref{eq157}) and their special cases like (\ref{eq155}), (\ref{eq158}) have respectively the following explicitly non-Euclidean (if $n>1$) forms: 
\begin{eqnarray}
\mathcal{G}&=&\mathcal{L}^{j}{}_{n}{}^{l}{}_{m}\varphi^{-1n}{}_{i}
\varphi^{-1m}{}_{k}d\varphi^{i}{}_{j}\otimes d\varphi^{k}{}_{l},\label{eq161}\\
\mathcal{G}&=&\varphi^{-1j}{}_{n}\varphi^{-1l}{}_{m}
\mathcal{R}^{n}{}_{i}{}^{m}{}_{k}d\varphi^{i}{}_{j}\otimes d\varphi^{k}{}_{l},\label{eq162}\\
\mathcal{G}&=&\left(A\varphi^{-1l}{}_{i}\varphi^{-1j}{}_{k}+
B\varphi^{-1j}{}_{i}\varphi^{-1l}{}_{k}\right)
d\varphi^{i}{}_{j}\otimes d\varphi^{k}{}_{l},\label{eq163}\\
\mathcal{G}&=&I_{1}\varphi^{-1j}{}_{m}
\varphi^{-1l}{}_{n}g^{mn}g_{ik}d\varphi^{i}{}_{j}\otimes d\varphi^{k}{}_{l}
+I_{2}g_{mn}\varphi^{-1m}{}_{i}\varphi^{-1n}{}_{k}g^{jl}
d\varphi^{i}{}_{j}\otimes d\varphi^{k}{}_{l}\nonumber\\
&+&\left(A\varphi^{-1l}{}_{i}\varphi^{-1j}{}_{k}+
B\varphi^{-1j}{}_{i}\varphi^{-1l}{}_{k}\right)
d\varphi^{i}{}_{j}\otimes d\varphi^{k}{}_{l}.\label{eq164}
\end{eqnarray}
Similar expansions with respect to the basic terms $d\varphi^{i}{}_{A}\otimes d\varphi^{j}{}_{B}$ may be done in general, e.g., for (\ref{eq158}) when $U$, $V$ may be different linear spaces. The corresponding formulae are structurally like the above ones. The explicit expressions like (\ref{eq161})--(\ref{eq164}) are in spite of their rather technical nature interesting in themselves and give an alternative analytical insight into the structure of expressions.

The restriction of the above metrics/kinetic energies to submanifolds of ${\rm GL}(V)$ is analytically achieved by specifying $\varphi^{i}{}_{j}$ as functions of some parameters, i.e., generalized coordinates $q^{\mu}$, $\mu=1,\ldots,f$. Technically this implies that the differentials $d\varphi^{i}{}_{j}$ are specified as 
\begin{equation}\label{eq165}
d\varphi^{i}{}_{j}=\frac{\partial\varphi^{i}{}_{j}}{\partial q^{\mu}}dq^{\mu}.
\end{equation}
In applications the mentioned submanifolds of ${\rm GL}(V)$ are usually its subgroups, e.g., special orthogonal ${\rm SO}(V,g)$, special linear ${\rm SL}(V)$, etc. In formulae one uses Lie algebras of the mentioned subgroups as geometrically and technically convenient non-holonomic velocities. Usually the choice of some particular models of $T$, i.e., of $\mathcal{G}$, is motivated by the particular symmetry demands under left and right regular translations in ${\rm GL}(V)$ or in the corresponding subgroup $G\subset{\rm GL}(V)$. The highest symmetry corresponds to (\ref{eq163}).

\subsubsection{Manifolds of scalar products and their non-holonomic velocities}

Let us now consider another pair of mutually related spaces, namely, the dual pair; $V=U^{\ast}$, $U=V^{\ast}$. Linear mappings from $V$ to $V^{\ast}$ are twice covariant tensors $\alpha$, i.e., elements of $V^{\ast}\otimes V^{\ast}\simeq{\rm L}\left(V,V^{\ast}\right)$, represented by matrices $\alpha_{ij}$. Similarly, linear mappings from $V^{\ast}$ to $V$ are twice contravariant tensors $\beta$, i.e., elements of $V\otimes V\simeq{\rm L}\left(V^{\ast},V\right)$, analytically represented by matrices $\beta^{ij}$. In the case of isomorphisms there exists a natural canonical bijection of ${\rm LI}\left(V,V^{\ast}\right)$ onto ${\rm LI}\left(V^{\ast},V\right)$; namely, according to the standard rules, interrelated are objects $\alpha\in V^{\ast}\otimes V^{\ast}$, $\beta\in V\otimes V$ such that
\begin{equation}\label{eq166}
\alpha_{ik}\beta^{kj}=\delta_{i}{}^{j},\qquad \beta^{ik}\alpha_{kj}=\delta^{i}{}_{j}.
\end{equation}
One can write simply
\begin{equation}\label{eq167}
\beta=\alpha^{-1},\qquad \alpha=\beta^{-1}.
\end{equation}
Just as when dealing with geodetic Hamiltonian systems on ${\rm GL}(V)$, one considers now again the very special case of ${\rm LI}(U,V)$, however, with completely new physical and geometrical peculiarities.

For any time evolutions 
\begin{equation}\label{eq167a}
\mathbb{R}\ni t\mapsto \alpha(t)\in{\rm LI}\left(V,V^{\ast}\right),\qquad \mathbb{R}\ni t\mapsto \beta(t)\in{\rm LI}\left(V^{\ast},V\right)
\end{equation}
the $\Omega,\widehat{\Omega}$-objects are as well defined as for any general $\mathbb{R}\ni t\mapsto \varphi(t)\in{\rm LI}\left(U,V\right)$. However, certain new features and peculiarities appear. The corresponding objects for curves in ${\rm LI}\left(V,V^{\ast}\right)$ are denoted by $\Omega[\alpha]\in{\rm L}\left(V^{\ast}\right)\simeq V^{\ast}\otimes V$ and $\widehat{\Omega}[\alpha]\in{\rm L}\left(V\right)\simeq V\otimes V^{\ast}$; more explicitly one can write them as $\Omega\left(\alpha,\dot{\alpha}\right)$, $\widehat{\Omega}\left(\alpha,\dot{\alpha}\right)$. Analytically they are given by
\begin{equation}\label{eq168}
\Omega[\alpha]_{i}{}^{j}=\dot{\alpha}_{ik}\alpha^{-1kj},\qquad \widehat{\Omega}[\alpha]^{i}{}_{j}=\alpha^{-1ik}\dot{\alpha}_{kj}.
\end{equation}
Obviously, they are interrelated by the following formulae:
\begin{eqnarray}
\Omega[\alpha]_{i}{}^{j}&=&\alpha_{ik}\widehat{\Omega}[\alpha]^{k}{}_{l}
\alpha^{-1lj},\qquad \Omega[\alpha]=\alpha\widehat{\Omega}[\alpha]\alpha^{-1},\label{eq169a}\\
\widehat{\Omega}[\alpha]^{i}{}_{j}&=&\alpha^{-1ik}\Omega[\alpha]_{k}{}^{l}
\alpha_{lj},\qquad \widehat{\Omega}[\alpha]=\alpha^{-1}\Omega[\alpha]\alpha.\label{eq169b}
\end{eqnarray}
Similarly, for curves in ${\rm LI}\left(V^{\ast},V\right)$ we use the symbols $\Omega[\beta]\in{\rm L}\left(V\right)\simeq V\otimes V^{\ast}$, $\widehat{\Omega}[\beta]\in{\rm L}\left(V^{\ast}\right)\simeq V^{\ast}\otimes V$ or $\Omega\left(\beta,\dot{\beta}\right)$, $\widehat{\Omega}\left(\beta,\dot{\beta}\right)$. The corresponding analytical expressions are as follows:
\begin{equation}\label{eq170}
\Omega[\beta]^{i}{}_{j}=\dot{\beta}^{ik}\beta^{-1}{}_{kj},\qquad \widehat{\Omega}[\beta]_{i}{}^{j}=\beta^{-1}{}_{ik}\dot{\beta}^{kj}.
\end{equation}
In analogy to (\ref{eq169a}), (\ref{eq169b}) there exists the following obvious relationships:
\begin{eqnarray}
\Omega[\beta]^{i}{}_{j}&=&\beta^{ik}\widehat{\Omega}[\beta]_{k}{}^{l}
\beta^{-1}{}_{lj},\qquad \Omega[\beta]=\beta\widehat{\Omega}[\beta]\beta^{-1},\label{eq171a}\\
\widehat{\Omega}[\beta]_{i}{}^{j}&=&\beta^{-1}{}_{ik}\Omega[\beta]^{k}{}_{l}
\beta^{lj},\qquad \widehat{\Omega}[\beta]=\beta^{-1}\Omega[\beta]\beta.\label{eq171b}
\end{eqnarray}
One can easily show that
\begin{eqnarray}
&&\Omega\left[\alpha^{-1}\right]=-\widehat{\Omega}[\alpha]=
\alpha^{-1}\widehat{\Omega}\left[\alpha^{-1}\right]\alpha\in V\otimes V^{\ast},\label{eq172a}\\
&&\Omega\left[\beta^{-1}\right]=-\widehat{\Omega}[\beta]=
\beta^{-1}\widehat{\Omega}\left[\beta^{-1}\right]\beta\in V^{\ast}\otimes V.\label{eq172b}
\end{eqnarray}

We are dealing here with the special case of the general scheme of ${\rm LI}\left(U,V\right)$-models, therefore, the properties of $\alpha\in V^{\ast}\otimes V^{\ast}\simeq{\rm L}\left(V,V^{\ast}\right)$ or $\beta\in V\otimes V\simeq{\rm L}\left(V^{\ast},V\right)$ may be analysed from the point of view of some additional structures like metrics in $V$, $V^{\ast}$, in analogy to metrics $g\in V^{\ast}\otimes V^{\ast}$, $\eta\in U^{\ast}\otimes U^{\ast}$ in mechanics of affine bodies. In particular, one can consider the rigid motion in the sense of those metrics; the $\Omega,\widehat{\Omega}$-objects become then skew-symmetric with respect to introduced metrics, i.e., are interpretable as "angular velocities". Obviously, when dealing with the manifold of scalar products in $V$, it would be rather artificial and exotic to introduce two independent metrics in $V$ and $V^{\ast}$ (as we did in mechanics of affine bodies in ${\rm LI}\left(U,V\right)$); rather some metric in $V$ and its contravariant inverse in $V^{\ast}$ would be used.

\noindent{\bf Remark:} 
One point is important here. In mechanical theory of systems with affine degrees of freedom, in ${\rm LI}\left(U,V\right)$, ${\rm GL}\left(V\right)$ as configuration spaces, it was rather natural to discuss constrained motion along subgroups of ${\rm GL}\left(V\right)$ or some orbits of the left or right actions of subgroups of ${\rm GL}\left(V\right)$, ${\rm GL}\left(U\right)$ on ${\rm LI}\left(U,V\right)$. Nonholonomic velocities were then the elements of the corresponding Lie subalgebras. And one concentrated on the left or right (or both) invariant metrics (kinetic energies) on configuration submanifolds. Nothing like this is useful in applications of dynamical systems on the manifold of scalar products. For bilinear forms, i.e., elements of $V^{\ast}\otimes V^{\ast}$ or $V\otimes V$, the symmetry/antisymmetry is well defined without any reference to something like a once fixed absolute scalar product. Usually we deal with symmetric or antisymmetric scalar products, e.g., if $V$ is over reals, with (pseudo-)Euclidean spaces or (generally) with symplectic spaces. When $V$ is over the complex field $\mathbb{C}$, the special stress is laid on Hermitian scalar products. Obviously, being twice covariant (or twice contravariant) tensors, not mixed ones, such objects cannot be multiplied in spite of their analytical matrix form. Even if we identify $V$ with $V^{\ast}$ using some pre-fixed reference scalar product and so identify the mentioned forms with linear transformations, nothing like the subgroup structure survives because, as a rule, the subsets of symmetric, antisymmetric, or Hermitian matrices are not closed under multiplication. And as a rule, the above objects $\Omega$, $\widehat{\Omega}$ do not form Lie algebras. Nevertheless, they are well-defined mixed tensors in $V$ and enable one to construct invariant quadratic scalars and, more generally, the family of basic scalars homogeneous in derivatives (generalised velocities) in a complete analogy to Lie-algebraic Casimir invariants (\ref{eq150}).

\subsubsection{Canonical Riemann structures, i.e., kinetic energies on manifolds of scalar products. High-symmetry models}

So, we would like to fix now some useful (at least hopefully) models of "kinetic energy", i.e., Riemannian structure, on appropriate manifolds of scalar products. As usual, it is impossible to be at the same time very general and computationally effective. And as a rule, it is the special cases, first of all the ones with high symmetry, that has a chance to be physically viable. So, step by step one reduces the interest to (\ref{eq156}), (\ref{eq157}), later on to (\ref{eq158}), (\ref{eq159}), then to (\ref{eq155}) and first of all to its special case $B=0$ (it is clear that the square-term controlled by $B$ is a merely secondary correction).

Explicitly, the Riemannian structure on the manifold ${\rm Sym}\left(V^{\ast}\otimes V^{\ast}\right)\subset V^{\ast}\otimes V^{\ast}$ of symmetric scalar products on $V$, constructed by analogy with the ${\rm GL}(V)$-prescription (\ref{eq155}), has the form corresponding to the kinetic energy
\begin{equation}\label{eq173}
T=\frac{A}{2}{\rm Tr}\left(\Omega[\alpha]^{2}\right)+
\frac{B}{2}\left({\rm Tr}\ \Omega[\alpha]\right)^{2},
\end{equation}
i.e., analytically to
\begin{equation}\label{eq174}
T=\frac{A}{2}\Omega[\alpha]_{i}{}^{j}\Omega[\alpha]_{j}{}^{i}+
\frac{B}{2}\left(\Omega[\alpha]_{i}{}^{i}\right)^{2}.
\end{equation}
In analogy to (\ref{eq163}) and using the standard Riemann expressions, we write for the underlying metric tensor that
\begin{equation}\label{eq175}
\mathcal{G}=\left(A\alpha^{-1li}\alpha^{-1jk}+B\alpha^{-1ji}\alpha^{-1lk}
\right)d\alpha_{ij}\otimes d\alpha_{kl},
\end{equation}
notice however some essential differences between (\ref{eq175}) and (\ref{eq163}). Using the favourite physicists way of thinking, we have the metric element
\begin{equation}\label{eq176}
ds^{2}=\left(A\alpha^{-1li}\alpha^{-1jk}+B\alpha^{-1ji}\alpha^{-1lk}
\right)d\alpha_{ij}d\alpha_{kl},
\end{equation}
where the pairs $(ij)$, $(kl)$ are, roughly speaking, bi-indices; their ordering does not matter when we deal with the manifold ${\rm Sym}\left(V^{\ast}\otimes V^{\ast}\right)$ of symmetric forms on $V$.

It is clear that when the forms $\alpha$ are symmetric, the plenty of "aesthetic" changes of ordering of indices is possible. Something similar, although a bit different, may be done for manifolds of symplectic forms. Equation (\ref{eq175}) may be written in the following form:
\begin{equation}\label{eq177}
\mathcal{G}=\mathcal{G}^{ijkl}(\alpha)d\alpha_{ij}\otimes d\alpha_{kl},
\end{equation}
where, let us notice carefully,
\begin{equation}\label{eq178}
\mathcal{G}^{abcd}=\frac{A}{2}
\left(\alpha^{-1ac}\alpha^{-1bd}+\alpha^{-1bc}\alpha^{-1ad}
\right)+B\alpha^{-1ab}\alpha^{-1cd}.
\end{equation}
The above formula implies that $\mathcal{G}^{abcd}$ have all necessary symmetry properties to represent some Riemannian metric on the manifold of real (or complex-analytic) symmetric scalar products,
\begin{equation}\label{eq179}
\mathcal{G}^{abcd}=\mathcal{G}^{bacd}=\mathcal{G}^{abdc}=\mathcal{G}^{cdab}
\end{equation}
(strictly speaking, the symmetry under the simultaneous exchange of pair $(ab)$, $(cd)$ holds without the representation (\ref{eq178})).

Although not interesting for our purposes here, it is nevertheless interesting in itself to search for natural metrics on the manifold of all non-degenerate bilinear forms on $V$, not necessarily symmetric ones. They work as follows when evaluated on pairs of tangent vectors:
\begin{equation}\label{eq180}
\mathcal{G}^{abcd}u_{ab}v_{cd},
\end{equation}
where
\begin{eqnarray}\label{eq181}
\mathcal{G}^{abcd}&=&A\alpha^{-1ab}\alpha^{-1cd}+\frac{D}{2}
\left(\alpha^{-1ac}\alpha^{-1bd}+\alpha^{-1ca}\alpha^{-1db}\right)
\nonumber\\
&+&
B\alpha^{-1ad}\alpha^{-1cb}+\frac{E}{2}
\left(\alpha^{-1ad}\alpha^{-1bc}+\alpha^{-1da}\alpha^{-1cb}\right)
\nonumber\\
&+&
G\alpha^{-1da}\alpha^{-1bc}+\frac{F}{2}
\left(\alpha^{-1ca}\alpha^{-1bd}+\alpha^{-1ac}\alpha^{-1db}\right).
\end{eqnarray}
Obviously, in an exactly the same way as we did above, we can construct natural scalar products on manifolds of non-degenerate twice contravariant tensors, i.e., Riemann structures on ${\rm Sym}\left(V\otimes V\right)$. Thus, we use the $\beta$-tensors, and, e.g., instead of formulae (\ref{eq173})--(\ref{eq178}) we obtain respectively that
\begin{equation}\label{eq182}
T=\frac{A}{2}{\rm Tr}\left(\Omega[\beta]^{2}\right)+
\frac{B}{2}\left({\rm Tr}\ \Omega[\beta]\right)^{2},
\end{equation}
i.e., analytically
\begin{equation}\label{eq183}
T=\frac{A}{2}\Omega[\beta]^{i}{}_{j}\Omega[\beta]^{j}{}_{i}+
\frac{B}{2}\left(\Omega[\beta]^{i}{}_{i}\right)^{2}.
\end{equation}
The underlying metric has the following form:
\begin{equation}\label{eq184}
\mathcal{G}=\left(A\beta^{-1}_{li}\beta^{-1}_{jk}+B\beta^{-1}_{ji}\beta^{-1}_{lk}
\right)d\beta^{ij}\otimes d\beta^{kl},
\end{equation}
in a full analogy to (\ref{eq175}). And just like in (\ref{eq177}), (\ref{eq178}) we have that
\begin{eqnarray}
\mathcal{G}&=&\mathcal{G}_{ijkl}\left(\beta\right)
d\beta^{ij}\otimes d\beta^{kl},\label{eq185}\\
\mathcal{G}_{abcd}&=&\frac{A}{2}
\left(\beta^{-1}_{ac}\beta^{-1}_{bd}+\beta^{-1}_{bc}\beta^{-1}_{ad}
\right)+B\beta^{-1}_{ab}\beta^{-1}_{cd}.\label{eq186}
\end{eqnarray}
All these metrics on the manifolds of scalar products are evidently curved, just like the Killing tensors on semisimple Lie groups.

\noindent{\bf Remark:} 
Let us notice that if some fixed metric $G\in {\rm Sym}\left(V^{\ast}\otimes V^{\ast}\right)$ is distinguished, it does not matter why, then the analogues of (\ref{eq158}), (\ref{eq159}), (\ref{eq164}) are also well defined both on ${\rm Sym}\left(V^{\ast}\otimes V^{\ast}\right)$ and ${\rm Sym}\left(V\otimes V\right)$. Analytically, the counterpart of (\ref{eq164}) as a metric on ${\rm Sym}\left(V^{\ast}\otimes V^{\ast}\right)$ is given by
\begin{eqnarray}\label{eq187}
\mathcal{G}&=&I_{1}\alpha^{-1jm}\alpha^{-1ln}G_{mn}G^{ik}d\alpha_{ij}\otimes d\alpha_{kl}
+
I_{2}\alpha^{-1im}\alpha^{-1kn}G_{mn}G^{jl}d\alpha_{ij}\otimes d\alpha_{kl}
\nonumber\\
&+&
\left(A\alpha^{-1li}\alpha^{-1jk}+B\alpha^{-1ji}\alpha^{-1lk}\right)
d\alpha_{ij}\otimes d\alpha_{kl},
\end{eqnarray}
where $I_{1}$, $I_{2}$, $A$, $B$ are some constants.

Obviously, $G$ with the upper-case indices is the contravariant inverse of $G$ with the lower-case indices, $G^{ik}G_{kj}=\delta^{i}{}_{j}$. One must be very careful when the inverse symbol is omitted for brevity in $\alpha^{-1ij}$, i.e., when we use the simplified notation $\alpha^{ij}$. This is again the inverse,
\begin{equation}\label{eq189}
\alpha^{ik}\alpha_{kj}=\delta^{i}{}_{j},
\end{equation}
however, one must remember that it is something else than the $G$-raising and lowering of indices,
\begin{equation}\label{eq190}
\alpha^{ij}=\alpha^{-1ij}\neq G^{ik}G^{jl}\alpha_{kl}.
\end{equation}

If nothing like some distinguished $G$ is fixed, we can consider only geodetic models, first of all homogeneous ones, characterized by high symmetries, like (\ref{eq173})--(\ref{eq175}). If for some physical reasons some certain reference metric $G$ is fixed, we have more possibilities. First of all, we can take the model (\ref{eq187}) of the kinetic energy and manipulate somehow with the constants $I_{1}$, $I_{2}$, $A$, $B$. But there are also some natural classes of potentials $V(\alpha)$ and the corresponding Lagrangians $L=T-V(\alpha)$. Namely, having at disposal two "metrics" $G,\alpha\in{\rm Sym}\left(V^{\ast}\otimes V^{\ast}\right)$ we can construct the mixed tensor $\widehat{\alpha}\in V\otimes V^{\ast}\simeq {\rm L}(V)$. Then in the $n$-dimensional space $V$ we can construct the system of $n$ independent basic scalars:
\begin{equation}\label{eq191}
I_{p}\left(\alpha,G\right):={\rm Tr}\left(\alpha^{p}\right),\qquad p=1,\ldots,n.
\end{equation}
According to the Cayley-Hamilton theorem, $I_{p}$ for any other integer $p$ may be expressed as a function of the above ones; again the property which was used in the study of deformation invariants. And the most natural and symmetric potentials $V$ are appropriately chosen functions of $I_{p}\left(\alpha,G\right)$.

Obviously, exactly the same may be done for systems on ${\rm Sym}\left(V\otimes V\right)$; there is no need to write down the obvious formulae. 

\subsubsection{Taking "translations" into account. Riemann structures on the manifolds of wave functions times scalar products}

Let us also mention another class of canonical Riemannian structures on the manifolds $V\times{\rm Sym}\left(V^{\ast}\otimes V^{\ast}\right)$. They have the following form:
\begin{equation}\label{eq192}
\mathcal{G}=M\alpha_{ij}du^{i}\otimes du^{j}+
\left(A\alpha^{-1li}\alpha^{-1jk}+B\alpha^{-1ji}\alpha^{-1lk}
\right)d\alpha_{ij}\otimes d\alpha_{kl},
\end{equation}
where $M$, $A$, $B$ are constants and $u^{i}$ are linear coordinates on $V$ corresponding to our choice of basis (therefore, they are simply elements of the dual basis in $V^{\ast}$). Kinetic energy based on (\ref{eq192}) is as follows:
\begin{equation}\label{eq193}
T=\frac{M}{2}\alpha_{ij}v^{i}v^{j}+
\frac{A}{2}{\rm Tr}\left(\Omega[\alpha]^{2}\right)+
\frac{B}{2}\left({\rm Tr}\ \Omega[\alpha]\right)^{2},
\end{equation}
where $v^{i}=du^{i}/dt$ is the "translational" velocity, in analogy to formulae for the affinely-rigid body. 

\noindent{\bf Remark:} 
Let us observe that in the "translational" part of $T$ the velocity $v^{i}$ is squared with the use of $\alpha$ itself, not with the use of some fixed metric $G\in{\rm Sym}\left(V^{\ast}\otimes V^{\ast}\right)$. This resembles some models of affine motion, where the translational velocity is
squared with the use of the Cauchy deformation tensor \cite{all_04,all_05}.

\subsection{Natural Lagrangians of high symmetry, dynamical\\ equations for scalar products and simple solutions}

Those were preliminary remarks based on intuitions developed during our earlier study of Hamiltonian systems on groups and homogeneous spaces, first of all on manifolds of affine and linear mappings (affinely-rigid bodies). Now we are well prepared to return to our proper subject, i.e., to the non-direct nonlinearity of hypothetical quantum mechanics. 

Let us go back to the complex linear space $W$ and consider the manifold of Hermitian scalar products there, i.e., the manifold of non-degenerate sesquilinear Hermitian forms, ${\rm Herm}\left(\overline{W}^{\ast}\otimes W^{\ast}\right)$. In principle they should be positively definite, but in many problems this restriction is not formally necessary. Analytically such scalar products $\Gamma$ are represented by Hermitian matrices $\left[\Gamma_{\overline{a}b}\right]$ in the sense that
\begin{equation}\label{eq194}
\Gamma(u,v)=\Gamma\left(u^{a}e_{a},v^{b}e_{b}\right)=
\Gamma_{\overline{a}b}\overline{u}^{\overline{a}}v^{b},
\end{equation}
where $e_{a}\in W$ are some basic vectors in $W$. 

Let ${\rm Herm}\left(\overline{W}^{\ast}\otimes W^{\ast}\right)$, or rather some its connected component (first of all the one consisting of positive forms), be our configuration space. Obviously, in spite of the complex character of $W$, the set ${\rm Herm}\left(\overline{W}^{\ast}\otimes W^{\ast}\right)$ is a real linear space, and the mentioned configuration space is a real manifold, an open subset of ${\rm Herm}\left(\overline{W}^{\ast}\otimes W^{\ast}\right)$. We are interested in Riemannian structures on this manifold and mainly in ones analogous to (\ref{eq173})--(\ref{eq175}) and (\ref{eq182})--(\ref{eq184}). To be more precise, those Riemannian structures will be restrictions of some Hermitian ones defined on the total $\overline{W}^{\ast}\otimes W^{\ast}$. We introduce them as some kinetic energy forms. 

\noindent{\bf Proposition:} 
Obviously, the only natural counterpart of (\ref{eq173})--(\ref{eq175}) is as follows:
\begin{equation}\label{eq195}
T=\frac{A}{2}\Gamma^{b\overline{c}}\Gamma^{d\overline{a}}
\dot{\Gamma}_{\overline{a}b}\dot{\Gamma}_{\overline{c}d}+
\frac{B}{2}\Gamma^{b\overline{a}}\Gamma^{d\overline{c}}
\dot{\Gamma}_{\overline{a}b}\dot{\Gamma}_{\overline{c}d},
\end{equation}
where $\Gamma$ with the upper-case indices is the contravariant inverse of one with the lower-case ones, $\Gamma^{-1}\in{\rm Herm}\left(W\otimes\overline{W}\right)$, i.e., analytically,
\begin{equation}\label{eq196}
\Gamma^{a\overline{c}}\Gamma_{\overline{c}b}=\delta^{a}{}_{b},\qquad
\Gamma_{\overline{a}b}\Gamma^{b\overline{c}}=
\delta_{\overline{a}}{}^{\overline{c}}.
\end{equation}

The metrics underlying (\ref{eq195}) are essentially curved and imply in a strong, non-perturba\-tive nonlinearity. The corresponding action functional will be denoted by
\begin{equation}\label{eq197}
I[\Gamma]=I[\Gamma,{\rm m}]+I[\Gamma,{\rm a}]=\int Tdt.
\end{equation}
Its terms controlled by constants $A$, $B$ were here denoted respectively by $I[\Gamma,{\rm m}]$, $I[\Gamma,{\rm a}]$; the labels ${\rm m}$ and ${\rm a}$ refer respectively to "main" and "additional". The reason is that evidently the $B$-term is a merely auxiliary correction and the model with $A=0$ would be meaningless. The $A$-term is a proper dynamics.

After some calculations one finds the following expressions for variational derivatives:
\begin{eqnarray}
\frac{\delta I[\Gamma,{\rm m}]}{\delta \Gamma_{\overline{a}b}(t)}&=&
-A\Gamma^{b\overline{n}}\left(\ddot{\Gamma}_{\overline{n}k}-
\dot{\Gamma}_{\overline{n}l}\Gamma^{l\overline{c}}\dot{\Gamma}_{\overline{c}k}
\right)\Gamma^{k\overline{a}},\label{eq198}\\
\frac{\delta I[\Gamma,{\rm a}]}{\delta \Gamma_{\overline{a}b}(t)}&=&
-B\Gamma^{l\overline{n}}\left(\ddot{\Gamma}_{\overline{n}l}-
\dot{\Gamma}_{\overline{n}k}\Gamma^{k\overline{c}}\dot{\Gamma}_{\overline{c}l}
\right)\Gamma^{b\overline{a}}.\label{eq199}
\end{eqnarray}
In analogy to doubly invariant Hamiltonian systems on semisimple Lie groups, one can show that the corresponding Euler-Lagrange equations
\begin{equation}\label{eq200}
\frac{\delta I[\Gamma]}{\delta \Gamma_{\overline{a}b}(t)}=0
\end{equation}
are solvable in terms of the matrix exponential function
\begin{equation}\label{eq201}
\Gamma_{\overline{r}s}(t)=G_{\overline{r}z}\exp(Et)^{z}{}_{s},
\end{equation}
where $E\in {\rm L}(W)\simeq W\otimes W^{\ast}$ and $G=\Gamma(0)\in{\rm Herm}\left(\overline{W}^{\ast}\otimes W^{\ast}\right)$ is the initial position in the configuration space ${\rm Herm}\left(\overline{W}^{\ast}\otimes W^{\ast}\right)$. It is clear that (\ref{eq201}) is a solution of (\ref{eq200}) for any $G$ and $E$. One can also easily show that $\Gamma(t)$ persists to be Hermitian for all $t\in\mathbb{R}$ if $E$ is $G$-Hermitian, i.e., if the sesquilinear form
\begin{equation}\label{eq202}
{}_{G}E_{\overline{r}s}:=G_{\overline{r}z}E^{z}{}_{s}
\end{equation}
is Hermitian.

This procedure is equivalent to the following one, based on the multiplication by matrix exponents on the left:
\begin{equation}\label{eq203}
\Gamma_{\overline{r}s}(t)=
\exp(Ft)_{\overline{r}}{}^{\overline{z}}G_{\overline{z}s},
\end{equation}
where again $G$ is arbitrary and $F\in {\rm L}\left(\overline{W}^{\ast}\right)\simeq \overline{W}^{\ast}\otimes \overline{W}$ must be $G$-Hermitian if $\Gamma$ is to be Hermitian for any $t\in\mathbb{R}$. In other words, the sesquilinear form
\begin{equation}\label{eq204}
\overline{W}^{\ast}\otimes W^{\ast}\ni\left(F_{G}\right)_{\overline{r}s}=
F_{\overline{r}}{}^{\overline{z}}G_{\overline{z}s}
\end{equation}
must be Hermitian.

It is seen that, depending on the choice of $E$ or $F$, $\Gamma$ may be oscillatory, exponentially growing or exponentially attenuating. All these situations may have something to do with decoherence, reduction, and other "paradoxes" of quantum mechanics.

\noindent{\bf Remark:} 
We were dealing here with the pure dynamics for the "scalar product" $\Gamma$, without any interaction with the "wave function" $\psi$. By the way, the dynamics for $\Gamma$ was purely amorphous in the sense that no fixed metric $G\in {\rm Herm}\left(\overline{W}^{\ast}\otimes W^{\ast}\right)$ was assumed. In principle we might assume some and admit for $T\left[\Gamma\right]$ something similar to (\ref{eq158}), (\ref{eq159}), (\ref{eq164}). No doubt, such expressions are definitely less convincing from the point of view of first principles of symmetry. On the other side, without using any fixed $G$, we have at our disposal only the above geodetic models (\ref{eq195}) for the pure dynamics of $\Gamma$. When some $G$ is distinguished, then just like in (\ref{eq191}) we can construct potentials invariantly built of some basic invariants of the following form:
\begin{equation}\label{eq205}
{\rm Tr}\left({}^{G}\Gamma^{p}\right),\qquad p=1,\ldots,n,
\end{equation}
where
\begin{equation}\label{eq206}
{}^{G}\Gamma^{r}{}_{s}:=G^{r\overline{z}}\Gamma_{\overline{z}s}.
\end{equation}
Incidentally, let us mention that instead of $I_{1},I_{2}$-controlled terms in (\ref{eq158}), (\ref{eq159}), (\ref{eq164}) one can try to use some simpler expressions based on the fixed $G$, just quadratic in generalized velocities with constant coefficients, although showing weaker symmetries, e.g.,
\begin{equation}\label{eq207}
\frac{I}{2}G^{b\overline{c}}G^{d\overline{a}}\dot{\Gamma}_{\overline{a}b}
\dot{\Gamma}_{\overline{c}d}+\frac{K}{2}
G^{b\overline{a}}G^{d\overline{c}}\dot{\Gamma}_{\overline{a}b}
\dot{\Gamma}_{\overline{c}d}.
\end{equation}

\noindent{\bf Proposition:} 
After introducing the new dynamical term (\ref{eq195}) (perhaps with some mentioned modifications), we must go back to all previous Lagrangians for $\psi$ and modify the corresponding variational derivatives on two levels:
\begin{enumerate}
    \item One of subsystems of equations of motion has the following form:
\begin{equation}\label{eq208}
\frac{\delta I\left[\Gamma,\psi\right]}{\delta \Gamma_{\overline{r}s}(t)}=0,
\end{equation}
where $I$ is built of the sum of all possible Lagrangians. We have just calculated (\ref{eq198}), (\ref{eq199}) for $I\left[\Gamma\right]$. But we must have (\ref{eq208}) just for the total $I\left[\Gamma,\psi\right]$ built of the total Lagrangian $L\left[\Gamma,\psi\right]$. Fortunately, in all previous Lagrangians $\Gamma$ enters in a purely algebraic way, so the calculations are relatively simple.

    \item For all previous terms one must revise the variational derivatives
\begin{equation}\label{eq209}
\frac{\delta I\left[\Gamma,\psi\right]}{\delta \psi^{a}}=0,\qquad 
\frac{\delta I\left[\Gamma,\psi\right]}{\delta \overline{\psi}^{\overline{a}}}=0
\end{equation}
taking, however, into account that $\Gamma$ is a dynamical quantity and the operation $d/dt$ in Euler-Lagrange expressions
\begin{equation}\label{eq210}
\frac{\partial L}{\partial \psi^{a}}-\frac{d}{dt}\frac{\partial L}{\partial \dot{\psi}^{a}},\qquad \frac{\partial L}{\partial \overline{\psi}^{\overline{a}}}-\frac{d}{dt}\frac{\partial L}{\partial \dot{\overline{\psi}}^{\overline{a}}}
\end{equation}
introduces some new terms involving $d\Gamma/dt$.
\end{enumerate}

\subsection{Total system of essentially nonlinear dynamical equations}

Let us review the corresponding equations of motion \cite{kyiv_08}. They are strongly nonlinear in $\left(\psi,\Gamma\right)$ and this nonlinearity is essential and non-perturbative. We have yet neither rigorous nor qualitative solutions, however the above remarks concerning the pure dynamics for $\Gamma$ seem to indicate that this kind of nonlinearity may be an alternative description of the open quantum system for $\psi$ with the surrounding symbolically represented by the dynamical $\Gamma$. The nonlinear interaction with $\Gamma$ might be perhaps a good candidate for explaining the aforementioned "paradoxes". This might be perhaps an approach alternative to that developed by Ingarden, Jamio\l kowski, Kossakowski \cite{IKO_97,Jam_72,Jam_74} and others.

\noindent{\bf Proposition:} 
Let us consider the most general Lagrangian:
\begin{eqnarray}\label{eq_8}
L&=&\alpha_{1}i\Gamma_{\bar{a}b}\left(\overline{\psi}{}^{\bar{a}}\dot{\psi}^{b}-
\dot{\overline{\psi}}{}^{\bar{a}}\psi^{b}\right)+
\alpha_{2}\Gamma_{\bar{a}b}\dot{\overline{\psi}}{}^{\bar{a}}
\dot{\psi}^{b}+
\left[\alpha_{4}\Gamma_{\bar{a}b}+\alpha_{5} H_{\bar{a}b}\right]
\overline{\psi}{}^{\bar{a}}\psi^{b}\nonumber\\
&+&\alpha_{3}\left[\Gamma^{b\bar{a}}+
\alpha_{9}\overline{\psi}{}^{\bar{a}}\psi^{b}\right]
\dot{\Gamma}_{\bar{a}b}+
\Omega[\psi,\Gamma]^{d\bar{c}b\bar{a}}
\dot{\Gamma}_{\bar{a}b}\dot{\Gamma}_{\bar{c}d}-V\left(\psi,\Gamma\right),
\end{eqnarray}
where
\begin{eqnarray}\label{eq_9}
\Omega[\psi,\Gamma]^{d\bar{c}b\bar{a}}&=&
\alpha_{6}\left[\Gamma^{d\bar{a}}+\alpha_{9}\overline{\psi}{}^{\bar{a}}\psi^{d}
\right]\left[\Gamma^{b\bar{c}}+\alpha_{9}\overline{\psi}{}^{\bar{c}}\psi^{b}
\right]+\alpha_{8}\overline{\psi}{}^{\bar{a}}\psi^{b}
\overline{\psi}{}^{\bar{c}}\psi^{d}
\nonumber\\
&+&
\alpha_{7}\left[\Gamma^{b\bar{a}}+\alpha_{9}\overline{\psi}{}^{\bar{a}}\psi^{b}
\right]\left[\Gamma^{d\bar{c}}+\alpha_{9}\overline{\psi}{}^{\bar{c}}\psi^{d}
\right]=\Omega[\psi,\Gamma]^{b\bar{a}d\bar{c}},
\end{eqnarray}
and the potential $V$ can be taken, for instance, in the following quartic form:
\begin{equation}\label{eq_10}
V\left(\psi,\Gamma\right)=\kappa\left(
\Gamma_{\bar{a}b}\overline{\psi}{}^{\bar{a}}\psi^{b}\right)^{2}.
\end{equation}

The first and second terms in (\ref{eq_8}) (those with 
$\alpha_{1}$ and $\alpha_{2}$) describe the free 
evolution of wave function $\psi$ while $\Gamma$ is fixed. 
The Lagrangian for trivial part of the linear dynamics 
(those with $\alpha_{4}$) can be also taken in the more general form 
$f\left(\Gamma_{\bar{a}b}\overline{\psi}{}^{\bar{a}}\psi^{b}\right)$, 
where $f:\mathbb{R}\rightarrow\mathbb{R}$.
The term with $\alpha_{5}$ corresponds to the Schr\"{o}dinger 
dynamics while $\Gamma$ is fixed and then
\begin{equation}\label{eq_14}
H^{a}{}_{b}=\Gamma^{a\bar{c}}H_{\bar{c}b}
\end{equation}
is the usual Hamilton operator. If we properly choose the 
constants $\alpha_{1}$ and $\alpha_{5}$, then we obtain precisely the Schr\"{o}dinger equation. 
The dynamics of the scalar product $\Gamma$ is described by the terms linear and quadratic in the time derivative of $\Gamma$.
In the above formulae $\overline{\psi}{}^{\bar{a}}=\overline{\psi^{a}}$ denotes the usual
complex conjugation and $\alpha_{i}$, $i=\overline{1,9}$, and $\kappa$ are some constants.

\noindent{\bf Remark:} 
The connection of the new constants $\alpha_{i}$ in (\ref{eq_8}) with the previous ones, i.e., $\alpha$, $\beta$, $\gamma$ (\ref{eq35}), $A$ and $B$ (\ref{eq195}), is as follows:
\begin{equation}\label{eq_14a}
\alpha_{1}=\alpha,\qquad \alpha_{2}=\beta,\qquad \alpha_{5}=-\gamma,\qquad \alpha_{6}=\frac{A}{2},\qquad \alpha_{7}=\frac{B}{2}.
\end{equation}

Applying the variational procedure we obtain the equations of motion as follows:
\begin{eqnarray}
\frac{\delta L}{\delta \overline{\psi}{}^{\bar{a}}}&=&
\alpha_{2}\Gamma_{\bar{a}b}\ddot{\psi}^{b}+\left(\alpha_{2}\dot{\Gamma}_{\bar{a}b}
-2\alpha_{1}i\Gamma_{\bar{a}b}\right)\dot{\psi}^{b}-
2\alpha_{8}\dot{\Gamma}_{\bar{a}b}\psi^{b}\dot{\Gamma}_{\bar{c}d}
\overline{\psi}{}^{\bar{c}}\psi^{d}
\nonumber\\
&-&2\alpha_{9}\left(\alpha_{6}\dot{\Gamma}_{\bar{a}d}\dot{\Gamma}_{\bar{c}b}+
\alpha_{7}\dot{\Gamma}_{\bar{a}b}\dot{\Gamma}_{\bar{c}d}\right)
\psi^{b}\left(\Gamma^{d\bar{c}}+\alpha_{9}\overline{\psi}{}^{\bar{c}}\psi^{d}\right)
\nonumber\\
&+&\left[\left(2\kappa \Gamma_{\bar{c}d}\overline{\psi}{}^{\bar{c}}\psi^{d}-
\alpha_{4}\right)\Gamma_{\bar{a}b}-\alpha_{5} H_{\bar{a}b}-
\left[\alpha_{3}\alpha_{9}+\alpha_{1}i\right]\dot{\Gamma}_{\bar{a}b}
\right]\psi^{b}=0\label{eq_19}
\end{eqnarray}
and
\begin{eqnarray}\label{eq_20}
\frac{\delta L}{\delta \Gamma_{\bar{a}b}}&=& 
2\Omega[\psi,\Gamma]^{b\bar{a}d\bar{c}}\ddot{\Gamma}_{\bar{c}d}+
2\dot{\Omega}[\psi,\Gamma]^{b\bar{a}d\bar{c}}\dot{\Gamma}_{\bar{c}d}+
\left(2\kappa \Gamma_{\bar{c}d}\overline{\psi}{}^{\bar{c}}\psi^{d}-
\alpha_{4}\right)\overline{\psi}{}^{\bar{a}}\psi^{b}\nonumber\\
&+&
2\Gamma^{d\bar{a}}\left[\alpha_{6}\Gamma^{b\bar{e}}\left(\Gamma^{f\bar{c}}+
\alpha_{9}\overline{\psi}{}^{\bar{c}}\psi^{f}\right)+
\alpha_{7}\Gamma^{b\bar{c}}\left(\Gamma^{f\bar{e}}+
\alpha_{9}\overline{\psi}{}^{\bar{e}}\psi^{f}\right)
\right]\dot{\Gamma}_{\bar{c}d}\dot{\Gamma}_{\bar{e}f}\nonumber\\
&-&
\alpha_{2}\dot{\overline{\psi}}{}^{\bar{a}}\dot{\psi}^{b}+
\left[\alpha_{3}\alpha_{9}+\alpha_{1}i\right]
\dot{\overline{\psi}}{}^{\bar{a}}\psi^{b}+
\left[\alpha_{3}\alpha_{9}-\alpha_{1}i\right]
\overline{\psi}{}^{\bar{a}}\dot{\psi}^{b}=0,
\end{eqnarray}
where
\begin{eqnarray}
\dot{\Omega}[\psi,\Gamma]^{b\bar{a}d\bar{c}}&=&
\alpha_{8}\left(\dot{\overline{\psi}}{}^{\bar{a}}\psi^{b}
\overline{\psi}{}^{\bar{c}}\psi^{d}+\overline{\psi}{}^{\bar{a}}\dot{\psi}^{b}
\overline{\psi}{}^{\bar{c}}\psi^{d}+\overline{\psi}{}^{\bar{a}}\psi^{b}
\dot{\overline{\psi}}{}^{\bar{c}}\psi^{d}+\overline{\psi}{}^{\bar{a}}\psi^{b}
\overline{\psi}{}^{\bar{c}}\dot{\psi}^{d}\right)
\nonumber\\
&+&
\alpha_{6}\alpha_{9}\left[\dot{\overline{\psi}}{}^{\bar{a}}\psi^{d}+
\overline{\psi}{}^{\bar{a}}\dot{\psi}^{d}\right]\left[\Gamma^{b\bar{c}}+
\alpha_{9}\overline{\psi}{}^{\bar{c}}\psi^{b}\right]\nonumber\\
&+&
\alpha_{6}\alpha_{9}\left[\dot{\overline{\psi}}{}^{\bar{c}}\psi^{b}+
\overline{\psi}{}^{\bar{c}}\dot{\psi}^{b}\right]\left[\Gamma^{d\bar{a}}+
\alpha_{9}\overline{\psi}{}^{\bar{a}}\psi^{d}\right]\nonumber\\
&+&
\alpha_{7}\alpha_{9}\left[\dot{\overline{\psi}}{}^{\bar{a}}\psi^{b}+
\overline{\psi}{}^{\bar{a}}\dot{\psi}^{b}\right]\left[\Gamma^{d\bar{c}}+
\alpha_{9}\overline{\psi}{}^{\bar{c}}\psi^{d}\right]\nonumber\\
&+&
\alpha_{7}\alpha_{9}\left[\dot{\overline{\psi}}{}^{\bar{c}}\psi^{d}+
\overline{\psi}{}^{\bar{c}}\dot{\psi}^{d}\right]\left[\Gamma^{b\bar{a}}+
\alpha_{9}\overline{\psi}{}^{\bar{a}}\psi^{b}\right]\label{eq_20a}\\
&-&
\alpha_{6}\left[\Gamma^{d\bar{e}}\Gamma^{f\bar{a}}\left(\Gamma^{b\bar{c}}+
\alpha_{9}\overline{\psi}{}^{\bar{c}}\psi^{b}\right)+
\Gamma^{b\bar{e}}\Gamma^{f\bar{c}}\left(\Gamma^{d\bar{a}}+
\alpha_{9}\overline{\psi}{}^{\bar{a}}\psi^{d}\right)\right]\dot{\Gamma}_{\bar{e}f} 
\nonumber\\
&-&
\alpha_{7}\left[\Gamma^{b\bar{e}}\Gamma^{f\bar{a}}\left(\Gamma^{d\bar{c}}+
\alpha_{9}\overline{\psi}{}^{\bar{c}}\psi^{d}\right)+
\Gamma^{d\bar{e}}\Gamma^{f\bar{c}}\left(\Gamma^{b\bar{a}}+
\alpha_{9}\overline{\psi}{}^{\bar{a}}\psi^{b}\right)\right]
\dot{\Gamma}_{\bar{e}f}.\nonumber
\end{eqnarray}
The Legendre transformation leads us to the following canonical variables:
\begin{eqnarray}
\pi_{b}&=&\frac{\partial L}{\partial \dot{\psi}^{b}}=
\alpha_{2}\Gamma_{\bar{a}b}\dot{\overline{\psi}}{}^{\bar{a}}+
\alpha_{1}i\Gamma_{\bar{a}b}\overline{\psi}{}^{\bar{a}},\label{eq_01a}\\
\overline{\pi}_{\bar{a}}&=&
\frac{\partial L}{\partial \dot{\overline{\psi}}{}^{\bar{a}}}=
\alpha_{2}\Gamma_{\bar{a}b}\dot{\psi}{}^{b}-
\alpha_{1}i\Gamma_{\bar{a}b}\psi^{b}, \label{eq_01b}\\
\pi^{\bar{a}b}&=&\frac{\partial L}{\partial \dot{\Gamma}_{\bar{a}b}}=
\alpha_{3}\left[\Gamma^{b\bar{a}}+
\alpha_{9}\overline{\psi}{}^{\bar{a}}\psi^{b}\right]+
2\Omega[\psi,\Gamma]^{b\bar{a}d\bar{c}}\dot{\Gamma}_{\bar{c}d}.\label{eq_02}
\end{eqnarray}
The energy of our $n$-level Hamiltonian system is as follows:
\begin{eqnarray}\label{eq_24}
E&=&\dot{\overline{\psi}}{}^{\bar{a}}\frac{\partial L}{\partial
\dot{\overline{\psi}}{}^{\bar{a}}}+\dot{\psi}^{b}\frac{\partial L}{\partial\dot{\psi}^{b}}+
\dot{\Gamma}_{\bar{a}b}\frac{\partial L}{\partial \dot{\Gamma}_{\bar{a}b}}-L=
-\left(\alpha_{4}\Gamma_{\bar{a}b}+\alpha_{5}
H_{\bar{a}b}\right)\overline{\psi}{}^{\bar{a}}\psi^{b}\nonumber\\
&+&
\alpha_{2}\Gamma_{\bar{a}b}\dot{\overline{\psi}}{}^{\bar{a}}
\dot{\psi}^{b}
+\Omega[\psi,\Gamma]^{\bar{a}b\bar{c}d}
\dot{\Gamma}_{\bar{a}b}\dot{\Gamma}_{\bar{c}d}
+\kappa\left(
\Gamma_{\bar{a}b}\overline{\psi}{}^{\bar{a}}\psi^{b}\right)^{2}.
\end{eqnarray}
Inverting the expressions (\ref{eq_01a}), (\ref{eq_01b}), (\ref{eq_02}) we obtain that
\begin{eqnarray}
\dot{\overline{\psi}}{}^{\bar{a}}&=&
\frac{1}{\alpha_{2}}\Gamma^{b\bar{a}}\pi_{b}-
\frac{\alpha_{1}}{\alpha_{2}}i\overline{\psi}{}^{\bar{a}},\qquad
\dot{\psi}{}^{b}=\frac{1}{\alpha_{2}}\Gamma^{b\bar{a}}\overline{\pi}_{\bar{a}}+
\frac{\alpha_{1}}{\alpha_{2}}i\psi^{b},\label{eq_03}\\
\dot{\Gamma}_{\bar{a}b}&=&
\frac{1}{2}\Omega[\psi,\Gamma]^{-1}_{\bar{a}b\bar{c}d}
\left(\pi^{\bar{c}d}-\alpha_{3}
\left[\Gamma^{d\bar{c}}+
\alpha_{9}\overline{\psi}{}^{\bar{c}}\psi^{d}\right]\right),\label{eq_04}
\end{eqnarray}
where
\begin{eqnarray}
\Omega[\psi,\Gamma]^{-1}_{\bar{a}b\bar{c}d}&=&
\Lambda[\psi,\Gamma]^{-1}_{\bar{a}b\bar{c}d}\nonumber\\
&-&
\frac{\alpha_{8}}{1+\alpha_{8}\theta_{2}[\psi,\Gamma]}
\Lambda[\psi,\Gamma]^{-1}_{\bar{a}b\bar{e}f}
\overline{\psi}{}^{\bar{e}}\psi^{f}
\Lambda[\psi,\Gamma]^{-1}_{\bar{c}d\bar{g}h}
\overline{\psi}{}^{\bar{g}}\psi^{h},\qquad \label{eq_04a1}\\
\Lambda[\psi,\Gamma]^{-1}_{\bar{a}b\bar{c}d}
&=&\frac{1}{\alpha_{6}}\lambda[\psi,\Gamma]^{-1}_{\bar{a}d}
\lambda[\psi,\Gamma]^{-1}_{\bar{c}b}\nonumber\\
&-&\frac{\alpha_{7}}{\alpha_{6}\left(
\alpha_{6}+n\alpha_{7}\right)}
\lambda[\psi,\Gamma]^{-1}_{\bar{a}b}\lambda[\psi,\Gamma]^{-1}_{\bar{c}d},
\label{eq_04a2}\\
\lambda[\psi,\Gamma]^{-1}_{\bar{a}b}&=&
\Gamma_{\bar{a}b}-\frac{\alpha_{9}}{1+\alpha_{9}\theta_{1}[\psi,\Gamma]}
\Gamma_{\bar{a}d}\Gamma_{\bar{c}b}\overline{\psi}{}^{\bar{c}}\psi^{d},
\label{eq_04a3}\\
\theta_{2}[\psi,\Gamma]&=&\Lambda[\psi,\Gamma]^{-1}_{\bar{a}b\bar{c}d}
\overline{\psi}{}^{\bar{a}}\psi^{b}
\overline{\psi}{}^{\bar{c}}\psi^{d}\nonumber\\
&=&\frac{
\alpha_{6}+\left(n-1\right)\alpha_{7}}{
\alpha_{6}\left(\alpha_{6}+n\alpha_{7}\right)}
\left(\frac{\theta_{1}[\psi,\Gamma]}{1+
\alpha_{9}\theta_{1}[\psi,\Gamma]}\right)^{2},\label{eq_04a4}\\
\theta_{1}[\psi,\Gamma]&=&\Gamma_{\bar{a}b}
\overline{\psi}{}^{\bar{a}}\psi^{b},\label{eq_04a5}
\end{eqnarray}
and then the Hamiltonian has the following form:
\begin{eqnarray}
H&=&\frac{1}{\alpha_{2}}\Gamma^{b\bar{a}}\overline{\pi}_{\bar{a}}\pi_{b}
+\frac{\alpha_{1}}{\alpha_{2}}i\left(\psi^{b}\pi_{\psi b}-
\overline{\psi}{}^{\bar{a}}\overline{\pi}_{\bar{a}}\right)
-\left[\left(\alpha_{4}-\frac{\alpha^{2}_{1}}{\alpha_{2}}\right)\Gamma_{\bar{a}b}
+\alpha_{5}H_{\bar{a}b}\right]
\overline{\psi}{}^{\bar{a}}\psi^{b}\nonumber\\
&+&\frac{1}{4}\Omega[\psi,\Gamma]^{-1}_{\bar{a}b\bar{c}d}\pi^{\bar{a}b}\pi^{\bar{c}d}
-\frac{\alpha_{3}}{2}\Omega[\psi,\Gamma]^{-1}_{\bar{a}b\bar{c}d}\left[\Gamma^{b\bar{a}}+
\alpha_{9}\overline{\psi}{}^{\bar{a}}\psi^{b}\right]\pi^{\bar{c}d}\nonumber\\
&+&\frac{\alpha^{2}_{3}}{4}
\Omega[\psi,\Gamma]^{-1}_{\bar{a}b\bar{c}d}\left[\Gamma^{b\bar{a}}+
\alpha_{9}\overline{\psi}{}^{\bar{a}}\psi^{b}\right]
\left[\Gamma^{d\bar{c}}+
\alpha_{9}\overline{\psi}{}^{\bar{c}}\psi^{d}\right]
+\kappa\left(
\Gamma_{\bar{a}b}\overline{\psi}{}^{\bar{a}}\psi^{b}\right)^{2}.\label{eq_05}
\end{eqnarray}

\noindent{\bf Remark:} 
Let us notice that if we suppose that the scalar product $\Gamma$ is
fixed, i.e.,  the equations of motion are as follows:
\begin{equation}\label{eq_33}
\alpha_{2}\ddot{\psi}^{a}-2\alpha_{1}i\dot{\psi}^{a}+
\left(2\kappa\theta_{1}\left[\psi,\Gamma\right]-
\alpha_{4}\right)\psi^{a}-\alpha_{5}H^{a}{}_{b}\psi^{b}=0,
\end{equation}
then taking all constants of the model to be equal to $0$ except of the following ones:
\begin{equation}\label{eq_34}
\alpha_{1}=\frac{\hbar}{2},\qquad \alpha_{5}=-1,
\end{equation}
we end up with the well-known usual Schr\"{o}dinger equation:
\begin{equation}\label{eq_36}
i\hbar\dot{\psi}^{a}=H^{a}{}_{b}\psi^{b}.
\end{equation}

\noindent{\bf Remark:} 
The first-order modified version of the Schr\"odinger equation is obtained when we suppose that $\Gamma$ is a dynamical variable
and $\alpha_{2}$ is equal to $0$, i.e.,
\begin{eqnarray}
i\hbar\dot{\psi}^{a}&=& H^{a}{}_{b}\psi^{b}-\left[\frac{i\hbar}{2}+\alpha_{3}\alpha_{9}\right]
\Gamma^{a\bar{c}}\dot{\Gamma}_{\bar{c}b}\psi^{b}\nonumber\\
&+&
\left(2\kappa\theta_{1}\left[\psi,\Gamma\right]-
\alpha_{4}\right)\psi^{a}-
2\alpha_{8}\Gamma^{a\bar{c}}\dot{\Gamma}_{\bar{c}b}\psi^{b}\dot{\Gamma}_{\bar{e}d}
\overline{\psi}{}^{\bar{e}}\psi^{d}\nonumber\\
&-&
2\alpha_{9}\Gamma^{a\bar{c}}\left(\alpha_{6}\dot{\Gamma}_{\bar{c}d}\dot{\Gamma}_{\bar{e}b}+
\alpha_{7}\dot{\Gamma}_{\bar{c}b}\dot{\Gamma}_{\bar{e}d}\right)
\psi^{b}\left(\Gamma^{d\bar{e}}+
\alpha_{9}\overline{\psi}{}^{\bar{e}}\psi^{d}\right),\label{eq_37}\\
2\Omega[\psi,\Gamma]^{b\bar{a}d\bar{c}}\ddot{\Gamma}_{\bar{c}d}
&=&\left[\frac{i\hbar}{2}-\alpha_{3}\alpha_{9}\right]
\overline{\psi}{}^{\bar{a}}\dot{\psi}^{b}-
\left[\frac{i\hbar}{2}+\alpha_{3}\alpha_{9}\right]
\dot{\overline{\psi}}{}^{\bar{a}}\psi^{b}\nonumber\\
&-&
2\alpha_{6}\Gamma^{d\bar{a}}\Gamma^{b\bar{e}}\left(\Gamma^{f\bar{c}}+
\alpha_{9}\overline{\psi}{}^{\bar{c}}\psi^{f}\right)
\dot{\Gamma}_{\bar{c}d}\dot{\Gamma}_{\bar{e}f}\nonumber\\
&-&
2\alpha_{7}\Gamma^{d\bar{a}}\Gamma^{b\bar{c}}\left(\Gamma^{f\bar{e}}+
\alpha_{9}\overline{\psi}{}^{\bar{e}}\psi^{f}\right)
\dot{\Gamma}_{\bar{c}d}\dot{\Gamma}_{\bar{e}f}\nonumber\\
&-&
\left(2\kappa\theta_{1}\left[\psi,\Gamma\right]-
\alpha_{4}\right)\overline{\psi}{}^{\bar{a}}\psi^{b}-
2\dot{\Omega}[\psi,\Gamma]^{b\bar{a}d\bar{c}}\dot{\Gamma}_{\bar{c}d}.
\label{eq_37a}
\end{eqnarray}

We can rewrite (\ref{eq_37}) in the following form:
\begin{equation}\label{eq_38}
i\hbar\dot{\psi}^{a}=H_{\rm eff}{}^{a}{}_{b}\psi^{b},
\end{equation}
where the effective Hamilton operator is given as follows: 
\begin{eqnarray}\label{eq_51}
H_{\rm eff}{}^{a}{}_{b}
&=&
H^{a}{}_{b}-\left[\frac{i\hbar}{2}+\alpha_{3}\alpha_{9}\right]
\Gamma^{a\bar{c}}\dot{\Gamma}_{\bar{c}b}\nonumber\\
&+&\left(2\kappa\theta_{1}\left[\psi,\Gamma\right]-
\alpha_{4}\right)\delta^{a}{}_{b}-
2\alpha_{8}\Gamma^{a\bar{c}}\dot{\Gamma}_{\bar{c}b}\dot{\Gamma}_{\bar{e}d}
\overline{\psi}{}^{\bar{e}}\psi^{d}\nonumber\\
&-&
2\alpha_{9}\Gamma^{a\bar{c}}\left(\alpha_{6}\dot{\Gamma}_{\bar{c}d}\dot{\Gamma}_{\bar{e}b}+
\alpha_{7}\dot{\Gamma}_{\bar{c}b}\dot{\Gamma}_{\bar{e}d}\right)\left(\Gamma^{d\bar{e}}+
\alpha_{9}\overline{\psi}{}^{\bar{e}}\psi^{d}\right).
\end{eqnarray}

\subsection{Invariance properties of our general Lagrangian}

So, if we investigate the invariance of our general Lagrangian (\ref{eq_8}) 
under the group ${\rm GL}(n,\mathbb{C})$ and consider some one-parameter group of transformations
\begin{equation}\label{eq_40}
\left\{\exp\left(A\tau\right):\tau\in\mathbb{R}\right\},\qquad A\in {\rm L}(n,\mathbb{C}),
\end{equation}
then the infinitesimal transformation rules for $\psi$ and $\Gamma$ are as follows:
\begin{equation}\label{eq_41}
\psi^{a}\mapsto L^{a}{}_{b}\psi^{b},\qquad
\Gamma^{a\bar{c}}\mapsto L^{a}{}_{b}\overline{L}{}^{\bar{c}}{}_{\bar{e}}\Gamma^{b\bar{e}},
\qquad
\Gamma_{\bar{a}b}\mapsto\Gamma_{\bar{c}d}
\overline{L^{-1}}{}^{\bar{c}}{}_{\bar{a}}L^{-1d}{}_{b},
\end{equation}
where
\begin{equation}\label{eq_42}
L^{a}{}_{b}=\delta^{a}{}_{b}+\epsilon A^{a}{}_{b},\qquad L^{-1a}{}_{b}\approx
\delta^{a}{}_{b}-\epsilon A^{a}{}_{b},\qquad \epsilon\approx 0.
\end{equation}
So, leaving only the first-order terms with respect to $\epsilon$ we obtain that the variations
of $\psi$ and $\Gamma$ are as follows:
\begin{eqnarray}
\delta\psi^{a}=\epsilon A^{a}{}_{b}\psi^{b},&\quad& \delta\overline{\psi}{}^{\bar{a}}=\epsilon
\overline{A}{}^{\bar{a}}{}_{\bar{c}}\overline{\psi}{}^{\bar{c}},\label{eq_43}\\
\delta \Gamma^{a\bar{c}}=\epsilon\left(A^{a}{}_{b}\Gamma^{b\bar{c}}+
\overline{A}{}^{\bar{c}}{}_{\bar{e}}\Gamma^{a\bar{e}}\right),&\quad& \delta
\Gamma_{\bar{a}b}=-\epsilon\left(\Gamma_{\bar{c}b}\overline{A}{}^{\bar{c}}{}_{\bar{a}}+
\Gamma_{\bar{a}d}A^{d}{}_{b}\right),\label{eq_43a}
\end{eqnarray}
then
\begin{eqnarray}
\frac{1}{\epsilon}\left(\frac{\partial L}{\partial \dot{\overline{\psi}}{}^{\bar{a}}}\delta
\overline{\psi}{}^{\bar{a}}+\frac{\partial L}{\partial
\dot{\psi}^{b}}\delta\psi^{b}\right)&=&\Gamma_{\bar{a}b}
\left(\alpha_{2}\dot{\overline{\psi}}{}^{\bar{a}}+
\alpha_{1}i\overline{\psi}{}^{\bar{a}}\right)A^{b}{}_{d}\psi^{d}
\nonumber\\
&+&\Gamma_{\bar{a}b}\left(\alpha_{2}\dot{\psi}^{b}-\alpha_{1}i\psi^{b}\right)
\overline{A}{}^{\bar{a}}{}_{\bar{c}}\overline{\psi}{}^{\bar{c}}\label{eq_44}
\end{eqnarray}
and
\begin{eqnarray}
\frac{1}{\epsilon}\frac{\partial L}{\partial\dot{\Gamma}_{\bar{a}b}}\delta
\Gamma_{\bar{a}b}&=&-\left[\alpha_{3}\left(\delta^{b}{}_{f}+
\alpha_{9}\Gamma_{\bar{a}f}\overline{\psi}{}^{\bar{a}}\psi^{b}\right)
+2\Omega\left[\psi,\Gamma\right]^{b\bar{a}d\bar{c}}
\Gamma_{\bar{a}f}\dot{\Gamma}_{\bar{c}d}\right]A^{f}{}_{b}\nonumber\\
&&-\left[\alpha_{3}\left(\delta^{\bar{a}}{}_{\bar{e}}+
\alpha_{9}\Gamma_{\bar{e}b}\overline{\psi}{}^{\bar{a}}\psi^{b}\right)
+2\Omega\left[\psi,\Gamma\right]^{b\bar{a}d\bar{c}}
\Gamma_{\bar{e}b}\dot{\Gamma}_{\bar{c}d}\right]
\overline{A}{}^{\bar{e}}{}_{\bar{a}}.\label{eq_45}
\end{eqnarray}
If we consider some fixed scalar product $\Gamma_{0}$ and take the $\Gamma_{0}$-hermitian $A$'s, then
\begin{equation}\label{eq_46}
A^{a}{}_{b}=\Gamma_{0}{}^{a\bar{c}}\widetilde{A}_{\bar{c}b},\qquad
\overline{A}^{\bar{a}}{}_{\bar{c}}=\widetilde{A}_{\bar{c}b}\Gamma_{0}^{b\bar{a}},\qquad \widetilde{A}{}^{\dag}=\widetilde{A},
\end{equation}
and therefore the above expressions can be written together in the matrix form:
\begin{equation}\label{eq_47}
\mathcal{J}\left(A\right)={\rm Tr}\left(V\widetilde{A}\right),
\end{equation}
where the hermitian tensor $V$ describing the system of conserved physical quantities is given by the following expression:
\begin{eqnarray}
V&=&\alpha_{2}\left(\psi\dot{\psi}^{\dag}\Gamma\Gamma^{-1}_{0}+\Gamma^{-1}_{0}\Gamma\dot{\psi}\psi^{\dag}\right)
+\left(\alpha_{1}i-\alpha_{3}\alpha_{9}\right)\psi\psi^{\dag}\Gamma\Gamma^{-1}_{0}
-2\alpha_{3}\Gamma^{-1}_{0}\nonumber\\
&-&
\left(\alpha_{1}i+\alpha_{3}\alpha_{9}\right)\Gamma^{-1}_{0}\Gamma\psi\psi^{\dag}
-2\left(\Gamma^{-1}_{0}\Gamma\omega\left[\psi,\Gamma\right]+\omega\left[\psi,\Gamma\right] \Gamma\Gamma^{-1}_{0}\right),\qquad \label{eq_48}
\end{eqnarray}
where
\begin{equation}\label{eq_48a}
\omega\left[\psi,\Gamma\right]^{b\bar{a}}=
\Omega\left[\psi,\Gamma\right]^{b\bar{a}d\bar{c}}\dot{\Gamma}_{\bar{c}d}.
\end{equation}
Similarly for the $\Gamma_{0}$-antihermitian $A$'s, i.e., when $\widetilde{A}^{\dag}=-\widetilde{A}$, we
obtain another hermitian tensor $W$ as a conserved value:
\begin{equation}\label{eq_49}
\mathcal{J}\left(A\right)={\rm Tr}\left(iW\widetilde{A}\right),
\end{equation}
where
\begin{eqnarray}
iW&=&\alpha_{2}\left(\psi\dot{\psi}^{\dag}\Gamma\Gamma^{-1}_{0}-
\Gamma^{-1}_{0}\Gamma\dot{\psi}\psi^{\dag}\right)
+\left(\alpha_{1}i-\alpha_{3}\alpha_{9}\right)
\psi\psi^{\dag}\Gamma\Gamma^{-1}_{0}\nonumber\\
&+&
\left(\alpha_{1}i+\alpha_{3}\alpha_{9}\right)\Gamma^{-1}_{0}\Gamma\psi\psi^{\dag}
+2\left(\Gamma^{-1}_{0}\Gamma\omega\left[\psi,\Gamma\right]-\omega\left[\psi,\Gamma\right] \Gamma\Gamma^{-1}_{0}\right).\label{eq_50}
\end{eqnarray}

\section*{Acknowledgements}

This paper contains results obtained within the framework of the research project 501 018 32/1992 financed from the Scientific Research Support Fund in 2007-2010. The authors are greatly indebted to the Ministry of Science and Higher Education for this financial support.

The authors are also very grateful to the referees for their valuable remarks and comments concerning this article.

\end{document}